\documentclass[aps, prb, showpacs, amsfonts, amsmath, amssymb, twocolumn, floatfix, superscriptaddress, 10pt]{revtex4-2}
\usepackage[final]{graphicx}
\usepackage{xcolor}
\usepackage[colorlinks=true, linkcolor=blue, citecolor=blue, urlcolor=blue]{hyperref}
\usepackage{verbatim}
\usepackage{csquotes}
\usepackage{braket}
\usepackage{mathtools}
\usepackage[normalem]{ulem}

\newcommand{\vct}[1]{\mathbf{#1}}

\renewcommand\Re{\operatorname{Re}}
\renewcommand\Im{\operatorname{Im}}
\newcommand\Tr{\operatorname{Tr}}
\allowdisplaybreaks

\begin{document}

% HEADER_begin ---------------------------------------------

\title{Heat radiation and transfer for nanoparticles in the presence of a cylinder}

\author{Kiryl Asheichyk}
\email[]{asheichyk@bsu.by}
\affiliation{Department of Theoretical Physics and Astrophysics, Belarusian State University, 5 Babruiskaya Street, 220006 Minsk, Belarus}   
\author{Matthias Kr{\"u}ger}
\email[]{matthias.kruger@uni-goettingen.de}
\affiliation{Institute for Theoretical Physics, Georg-August-Universit{\"a}t G{\"o}ttingen, 37073 G{\"o}ttingen, Germany}

\date{\today}

\begin{abstract}
We study heat radiation and radiative heat transfer for nanoparticles in the presence of an infinitely long cylinder in different geometrical configurations, based on its electromagnetic Green's tensor. The heat radiation of a single particle can be enhanced by placing it close to a nanowire, and this enhancement can be much larger as compared to  placing it close to a plate of the same material. The heat transfer along a cylinder decays much slower than through empty vacuum, being especially long-ranged in the case of a perfectly conducting nanowire, and showing nonmonotonic behavior in the case of a SiC cylinder. Exploring the dependence on the relative azimuthal angle of the nanoparticles, we find that the results are insensitive to small angles, but they can be drastically different when the angle is large, depending on the material. Finally, we demonstrate that a cylinder can either enhance or block the heat flux when placed perpendicular to the interparticle distance line, where the blocking in particular is strongly enhanced compared to the geometry of a sphere of same radius. 
\end{abstract}


\maketitle

% HEADER_end -----------------------------------------------

% MAIN_PART_begin ------------------------------------------

\section{Introduction}
\label{sec:Introduction}
Fundamental understanding of radiative heat exchange in complex micro- and nanosystems is of high practical importance for a variety of applications~\cite{Song2021, Biehs2021, Basu2007}. Over the past decade, this understanding has become possible thanks to the development of the corresponding theoretical frameworks~\cite{Bimonte2017,Song2021, Biehs2021}, where expressions for many-body heat radiation (HR) and radiative heat transfer (HT) are derived based on fluctuational electrodynamics~\cite{Rytov1958, Rytov1989}.

With these frameworks, numerous paradigmatic configurations were investigated, mainly concerning HT for a collection of small particles~\cite{Ben-Abdallah2011, Nikbakht2014, Choubdar2016, Dong2017_1, Phan2013, Nikbakht2017, Dong2017_2, Luo2019, Ben-Abdallah2013, Luo2021, Ordonez-Miranda2018, Kathmann2018, Tervo2020, Luo2020, Liu2023, Walter2022}, HT in many-body planar structures~\cite{Zheng2011, Messina2012, Messina2016, Song2018, Simchi2017, He2019_1, Latella2017, Latella2018, Kan2019}, and HT between two small particles in the presence of an arbitrarily sized object~\cite{Saaskilahti2014, Dong2018, Messina2018, Asheichyk2018, Zhang2019_1, Zhang2019_2, He2019_2, Fang2022, Asheichyk2017, Asheichyk2022, Wang2024, Asheichyk2023}. While the numerical feasibility of the first two cases becomes worse upon increasing the number of particles or planar structures, in the latter case, it depends on how complicated the considered object is. Therefore, the performed computations are mainly restricted to objects with an analytical scattering matrix, such that their Green's function (GF) is expressed as a sum and (or) integral of known functions. These are a plate~\cite{Saaskilahti2014, Dong2018, Messina2018, Asheichyk2018, Zhang2019_1, Zhang2019_2, He2019_2, Fang2022} or a two-plates cavity~\cite{Saaskilahti2014, Asheichyk2018, Asheichyk2023}, a sphere~\cite{Asheichyk2017} or a spherical cavity~\cite{Asheichyk2017, Asheichyk2018}, and a cylinder~\cite{Asheichyk2022, Wang2024} or a cylindrical cavity~\cite{Asheichyk2023}.  

Cylindrical objects in the context of HR and HT are well explored: HR of a cylinder alone~\cite{Kruger2011, Golyk2012, Rodriguez2012, Rodriguez2013_1, Ohman1961, Bimonte2009, Fan2009, Singer2011} as well as HT between a cylinder and another object~\cite{Rodriguez2012, Rodriguez2013_1, McCauley2012, Rodriguez2013_2, Xiao2023} were investigated both theoretically~\cite{Kruger2011, Golyk2012, Rodriguez2012, Rodriguez2013_1, McCauley2012, Rodriguez2013_2, Xiao2023} and experimentally~\cite{Ohman1961, Bimonte2009, Fan2009, Singer2011} by several research groups. However, HT between two particles in the presence of a cylinder was studied only very recently~\cite{Asheichyk2022, Wang2024}. In Ref.~\cite{Asheichyk2022}, we showed that a well-conducting cylinder acts as an excellent waveguide, transferring energy to large distances much more efficiently than planar or spherical objects.

In this work, we discover further effects that a cylinder has on  HT and HR of nanoparticles, extending the studies of Ref.~\cite{Asheichyk2022} to a larger set of geometrical configurations and materials. In particular, we show that the energy flux along a dielectric cylinder is several orders of magnitude larger than the vacuum flux, even greatly surpassing the transfer along a well-conducting cylinder for a certain range of parameters. Rotation of one particle around a cylinder relative to the other (by keeping the interparticle distance roughly unchanged) has a large or small effect depending on the material. When the particles are placed perpendicular to the cylinder axis, the flux can be either enhanced or blocked. In addition, we demonstrate that a particle placed close to a cylinder radiates (or, in other words, cools down) much stronger than when it is in isolation, and also much stronger compared to placing it close a plate. 

The paper is organized as follows. In Sec.~\ref{sec:HR}, we study the HR of a particle in the presence of a cylinder. Section~\ref{sec:HT_parallel} investigates the HT between two particles placed parallel to the cylinder axis, whereas Sec.~\ref{sec:HT_perpendicular} is devoted to the perpendicular configuration. We close the main part with a summary and discussion in Sec.~\ref{sec:Conclusion}. In the Appendix, we study the GF of a cylinder in detail, giving its expressions for various cases.

\section{Heat radiation}
\label{sec:HR}
\begin{figure}[!t]
\begin{center}
\includegraphics[width=0.95\linewidth]{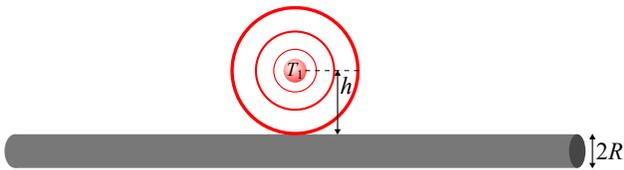}
\end{center}
\caption{\label{fig:SystemHR}Heat radiation of particle $ 1 $ at temperature $ T_1 $ in the presence of an infinitely long cylinder of radius $ R $. When the distance $ h $ is small (i.e., near-field regime), the cylinder strongly affects the particle's radiation (see Fig.~\ref{fig:HR}).}
\end{figure}

Consider particle $ 1 $ at temperature $ T_1 $ placed close to a cylinder of radius $ R $ at distance $ h $ from its surface, as depicted in Fig.~\ref{fig:SystemHR}. The cylinder is assumed to be infinitely long, such that its scattering matrix is known analytically (see the Appendix). We aim to compute HR of the particle in this system, i.e., the rate of heat emitted by the particle, taking into account possible reabsorption. To simplify computations, we use the dipolar point particle (PP) limit, such that particle $ 1 $ is small compared to the thermal wavelength, the particle's skin depth, and the distance $ h $~\cite{Asheichyk2017}. In addition, the limit implies that magnetic response of the particle can be neglected (i.e., its magnetic permeability $ \mu = 1 $). With these conditions, the multiple scatterings from the particle are neglected, and it can be modeled as an electrical dipole~\cite{Asheichyk2017}. HR then reads~\cite{Asheichyk2017}
\begin{equation}
H_1^{(1)} = \frac{8\hbar}{c^2}\int_0^\infty d\omega \frac{\omega^3}{e^{\frac{\hbar\omega}{k_{\textrm{B}}T_1}}-1}\Im(\alpha_1)\Tr\Im\mathbb{G}(\vct{r}_1, \vct{r}_1).
\label{eq:HR}
\end{equation}
Here, $ \Tr\Im\mathbb{G}(\vct{r}_1, \vct{r}_1) $ is the trace of the imaginary part of the dyadic Green's function (GF) of the cylinder, evaluated at the particle's position $ \vct{r}_1 $, which can be found in the Appendix [Eqs.~\eqref{eq:GT_trace} and~\eqref{eq:GT_trace_mirror}]. This trace encodes the geometry of the system, and thus determines how the cylinder affects the HR. Note that $ \mathbb{G} $ is also a function of $ \omega $. $ c $ is the speed of light in vacuum, and $ \hbar $ and $ k_{\textrm{B}} $ are Planck's and Boltzmann's constants, respectively. $ \alpha_1 $ is the particle's polarizability, characterizing its natural radiation (or absorption) strength,
\begin{equation}
\alpha_1 (\omega)= \frac{\varepsilon_1(\omega)-1}{\varepsilon_1(\omega)+2}R_1^3,
\label{eq:polarizability}
\end{equation}
with $ R_1 $ and $ \varepsilon_1(\omega) $ being the radius and the frequency-dependent dielectric permittivity of the particle, respectively. Given Eq.~\eqref{eq:polarizability}, $ H_1^{(1)} $ is proportional to the particle's volume $ V_1 $. 

For demonstration of the HR, we choose a SiC (alpha silicon carbide) particle at temperature $ T_1 = 300 \ \textrm{K} $, with the following permittivity~\cite{Spitzer1959}:
\begin{equation}
\varepsilon_1(\omega) = \varepsilon_{\textrm{SiC}}(\omega) = \varepsilon_\infty\frac{\omega^2-\omega_{\textrm{LO}}^2+i\omega\gamma}{\omega^2-\omega_{\textrm{TO}}^2+i\omega\gamma},
\label{eq:epsilon_SiC}
\end{equation}
with $ \varepsilon_\infty=6.7 $, $ \omega_{\textrm{LO}}=1.82\times10^{14} \ \textrm{rad} \ \textrm{s}^{-1} $, $ \omega_{\textrm{TO}}=1.49\times10^{14} \ \textrm{rad} \ \textrm{s}^{-1} $, $ \gamma=8.93\times10^{11} \ \textrm{rad} \ \textrm{s}^{-1} $. 
The corresponding thermal wavelength (which sets the short-wavelength cutoff in the radiation spectrum, and around which the dominantly contributing wavelengths are concentrated) is $ \lambda_{T_1} = \frac{\hbar c}{k_{\textrm{B}}T_1} = 7.63 \times 10^{-6} \ \textrm{m} $; the corresponding dominant frequency $ \omega_0 = 1.75 \times 10^{14} \ \textrm{rad} \ \textrm{s}^{-1} $, giving the maximum of the radiation spectrum, is the resonance frequency of $ \alpha_{\textrm{SiC}} $. For the cylinder, we consider three different materials -- SiC, gold, and a perfect conductor -- neglecting the magnetic response, i.e., $ \mu = 1 $ (note that the GF given in the Appendix can be used for arbitrary $ \mu $). The dielectric response of a SiC cylinder is modeled by $ \varepsilon_{\textrm{SiC}}(\omega) $ in Eq.~\eqref{eq:epsilon_SiC}, while for a gold cylinder, the Drude model is used~\cite{Ordal1983},
\begin{equation}
\varepsilon_{\rm Au}(\omega) = 1-\frac{\omega_{\textrm{p}}^2}{\omega(\omega+i\omega_{\tau})},
\label{eq:epsilon_gold}
\end{equation}
with $ \omega_{\textrm{p}} = 1.37\times 10^{16} \ {\rm rad} \ {\rm s}^{-1} $ and $ \omega_{\tau} = 4.06\times 10^{13} \ {\rm rad} \ {\rm s}^{-1} $. The perfect conductor is modeled using the corresponding scattering matrices in Eqs.~\eqref{eq:TMMmirror}--\eqref{eq:TMNmirror} (formally, it is the $ |\varepsilon| \to \infty $ limit).

\begin{figure*}[!t]
\begin{center}
\includegraphics[width=1.0\linewidth]{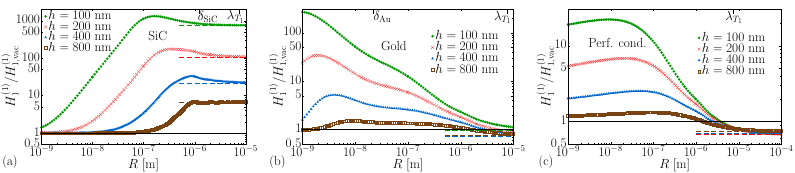}
\end{center}
\caption{\label{fig:HR}Heat radiation of a SiC particle at temperature $ T_1 = 300 \ \textrm{K} $ in the presence of a cylinder, as a function of the cylinder radius $ R $, normalized by the heat radiation in isolation. The results are given for different materials of the cylinder [(a) SiC, (b) gold, (c) perfect conductor] and for different distances $ h $ from the particle to the cylinder surface (see Fig.~\ref{fig:SystemHR}). Dashed lines show the heat radiation in the presence of a plate of the corresponding material for the corresponding $ h $. $ \lambda_{T_1} = 7.63 \times 10^{-6} \ \textrm{m} $ is the thermal wavelength, while $ \delta_{\textrm{SiC}} = 1.2 \times 10^{-6} \ \textrm{m} $ and $ \delta_{\textrm{Au}} = 2.2 \times 10^{-8} \ \textrm{m} $ are the skin depths of SiC and gold, respectively. Solid line at a value of $ 1 $ is included as a guide to the eye.}
\end{figure*}

The HR in the presence of a cylinder, normalized by the HR of an isolated particle, $ H_{1,\textrm{vac}}^{(1)} $ (see Eq.~(22) in Ref.~\citep{Asheichyk2017}), is given in Fig.~\ref{fig:HR} as a function of $ R $ for different materials of the cylinder and near-field distances $ h $. Overall, we can see that a cylinder largely amplifies the HR, with the effect becoming stronger as $ h $ decreases. For all considered materials, the ratio $ H_{1}^{(1)}/H_{1,\textrm{vac}}^{(1)} $ is a nonmonotonic function of $ R $, such that there is an optimal radius giving the maximal amplification (the maximum for $ h = 100 \ \textrm{nm} $ in Fig.~\ref{fig:HR}(b) is not evident, as it appears at the smallest considered $ R = 1 \ \textrm{nm} $). This radius shifts to smaller values with a decrease of $ h $.

\begin{figure}[!b]
\begin{center}
\includegraphics[width=1.0\linewidth]{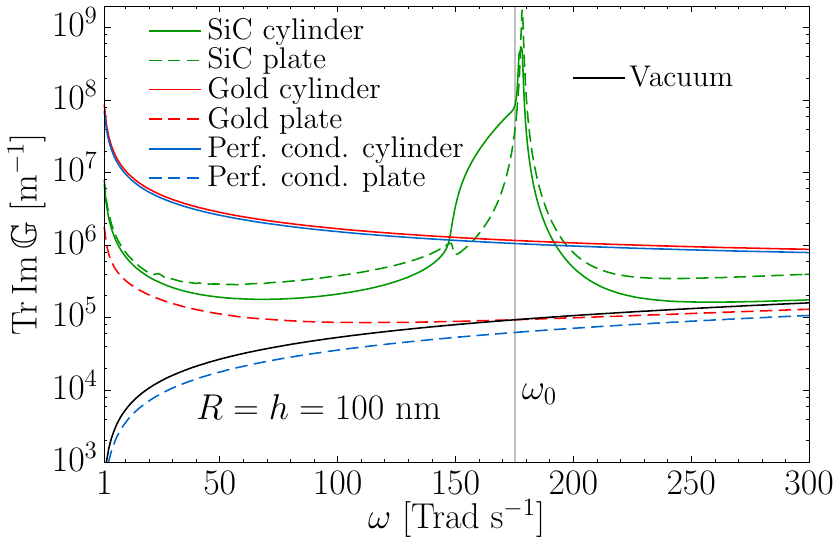}
\end{center}
\caption{\label{fig:SpecHR}$ \Tr\Im\mathbb{G}(\vct{r}_1, \vct{r}_1) $ as a function of frequency as appearing  in the radiation spectrum of a particle in Eq.~\eqref{eq:HR}. Here, we consider a cylinder ($ R = h = 100 \ \textrm{nm} $), plate (same $ h $), or no object (vacuum). The vertical line shows the resonance frequency $ \omega_0 $ of the SiC particle polarizability, which is the dominant frequency in heat radiation [see Eqs.~\eqref{eq:HR}--\eqref{eq:epsilon_SiC}].}
\end{figure}

The strongest enhancement of the HR is observed for a SiC cylinder [shown in Fig.~\ref{fig:HR}(a)], which can be attributed to similar resonances of the spectra for $ \alpha_1 $ and $ \Tr\Im\mathbb{G} $ (see Fig.~\ref{fig:SpecHR}). When $ R $ is small, there is a little effect, as the cylinder becomes transparent. With growth of $ R $, the HR strongly increases, reaching its maximum when $ R $ becomes larger than $ h $. For $ h = 800 \ \textrm{nm} $, the maximal amplification is around $ 7 $, while it exceeds $ 1300 $ for $ h = 100 \ \textrm{nm} $. When $ R $ becomes comparable to $ \lambda_{T_1} $, the result converges to the HR in the presence of a plate. The maximal ratio between the HR for cylinder and plate geometries grows with a decrease of $ h $, but it does not exceed $ 2 $ for the considered $ h \geq 100 \ \textrm{nm} $.

In contrast to SiC, the spectrum of the gold cylinder GF has overall a smaller amplitude and no peaks (see Fig.~\ref{fig:SpecHR}), such that $ H_{1}^{(1)}/H_{1,\textrm{vac}}^{(1)} $ is also smaller [see Fig.~\ref{fig:HR}(b)], with its maximum varying between $ 1.5 $ (for $ h = 800 \ \textrm{nm} $) and $ 264 $ (for $ h = 100 \ \textrm{nm} $). Compared to SiC, this maximum is reached at a much smaller $ R $ (e.g., $ R_{\textrm{max}} \approx 1 \ \textrm{nm} $ for $ h = 100 \ \textrm{nm} $), which we attribute to a much smaller skin depth ($ \delta_{\textrm{Au}} = 2.2 \times 10^{-8} \ \textrm{m} $ versus $ \delta_{\textrm{SiC}} = 1.2 \times 10^{-6} \ \textrm{m} $). The plate limit for gold is below the vacuum HR, such that a thin gold cylinder can outperform a gold plate by several orders of magnitude.

A perfectly conducting cylinder shows a smaller amplification compared to SiC and gold, with its maximum being $ 22 $ for $ h = 100 \ \textrm{nm} $ [see Fig.~\ref{fig:HR}(c)]. Notably, the convergence to the vacuum HR is very slow, such that $ H_{1}^{(1)}/H_{1,\textrm{vac}}^{(1)} $ for a thin cylinder is significantly above $ 1 $ even for the largest considered $ h = 800 \ \textrm{nm} $; in other words, when SiC and gold cylinders become transparent, a perfectly conducting one still affects the particle's emission. As in the case of gold, the plate limit is below the vacuum one, leading to a large cylinder-plate amplification factor. Importantly, for a perfectly conducting plate, one can apply the method of images, and only one image dipole is required.  $ \Tr\Im\mathbb{G} $ is known analytically, leading to an amplification factor between  $ 2/3 $ and $ 2 $ compared to the case of a free particle~\cite{Asheichyk2017, Tai1994}, indicating that the amplification may be limited by the number of image dipoles. For a cylinder, the method of images is more complicated, involving multiple image dipoles. This larger number of image dipoles might be the reason why a cylinder allows a larger amplification as seen in Fig.~\ref{fig:HR}(c). It is worth noticing that a perfectly conducting cylinder requires many fewer multipoles [indexed by $ n $ in Eq.~\eqref{eq:GT_trace_mirror}] for the convergence of $ \Tr\Im\mathbb{G} $ compared to SiC and gold [for which Eq.~\eqref{eq:GT_trace} is used]: For example, $ n_{\textrm{max}} \approx 5 $ is enough for a perfect conductor when $ h = 100 \ \textrm{nm} $ and $ R = 1 \ \mu\textrm{m} $, whereas $ n_{\textrm{max}} \approx 50 $ is required for the same $ h $ and $ R $ in the case of SiC or gold.

The result for the perfect conductor is remarkable: In this case, the cylinder strictly absorbs no energy, and the energy emitted in the HR of the particle must travel to infinity (in contrast to gold and SiC, where the energy can be absorbed by the cylinder). In this regard, the difference to the perfectly conducting plate is worth noting: The nanoparticle seems to excite waves traveling along the cylinder, but it seems not to be able to excite (noticeable) waves traveling along the plate. Interestingly, the relative factor becomes even larger for smaller frequencies (or larger wavelengths): As shown in Fig.~\ref{fig:SpecHR}, with a decrease of $ \omega $, $ \Tr\Im\mathbb{G} $ grows for a perfectly conducting cylinder, but it decays for a plate. This means that the effect of a cylinder should be most pronounced for small temperature $ T_1 $ (large $ \lambda_{T_1} $).

\section{Heat transfer: parallel configuration}
\label{sec:HT_parallel}
Let us now turn to the HT from particle $ 1 $ at temperature $ T_1 $ to particle $ 2 $ in the presence of a cylinder. As in Sec.~\ref{sec:HR}, we use the PP limit (here, for both particles), which requires that the radius of each particle is much smaller than the distance between them (in addition to the conditions discussed in Sec.~\ref{sec:HR}). In this limit, the HT reads~\cite{Asheichyk2017, Ben-Abdallah2011}
\begin{align}
\notag H_1^{(2)} = \ & \frac{32\pi\hbar}{c^4} \int_0^\infty d\omega \frac{\omega^5}{e^{\frac{\hbar\omega}{k_{\textrm{B}}T_1}}-1}\Im(\alpha_1)\Im(\alpha_2)\\
& \times \Tr\left[\mathbb{G}(\vct{r}_1, \vct{r}_2)\mathbb{G}^{\dagger}(\vct{r}_1, \vct{r}_2)\right],
\label{eq:HT}
\end{align}
where $ \Tr\left[\mathbb{G}(\vct{r}_1, \vct{r}_2)\mathbb{G}^{\dagger}(\vct{r}_1, \vct{r}_2)\right] $ is the trace of the matrix product of the GF of the cylinder (evaluated at the particles' positions $ \vct{r}_1 $ and $ \vct{r}_2 $) and its conjugate transpose. The polarizabilities $ \alpha_1 $ and $ \alpha_2 $ are given by Eq.~\eqref{eq:polarizability} (with the corresponding particle index), such that the HT is proportional to volumes of the particles $ V_1 $ and $ V_2 $. We thus do not give the particles' radii explicitly but instead normalize the HT by $ V_1V_2 $.

\begin{figure}[!t]
\begin{center}
\includegraphics[width=0.95\linewidth]{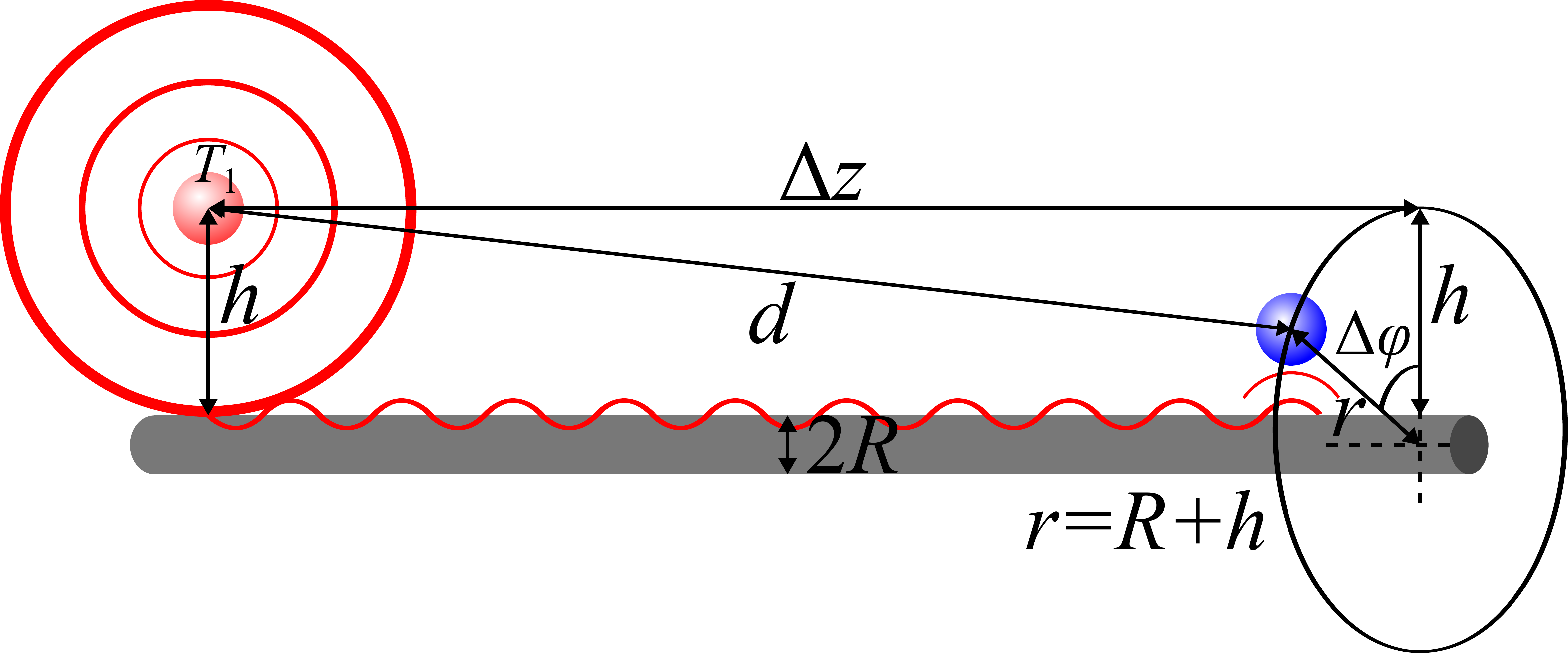}
\end{center}
\caption{\label{fig:SystemHTpar}Radiative heat transfer from particle $ 1 $ at temperature $ T_1 $ to particle $ 2 $, placed close to an infinitely long cylinder parallel ($\Delta\varphi=0 $) to the cylinder axis. The near-field energy radiated by the first particle is captured by the cylinder and guided in the needed direction to the second particle. Such a directional configuration leads to a highly efficient heat transfer even for far-separated particles (see Fig.~\ref{fig:HTpar}). Variation of relative azimuthal angle $ \Delta\varphi $ (provided that $ d \approx \Delta z $) can lead to a mild or strong change of the heat transfer, depending on the material of the cylinder (see insets of Fig.~\ref{fig:HTpar}).}
\end{figure}

\begin{figure*}[!t]
\begin{center}
\includegraphics[width=1.0\linewidth]{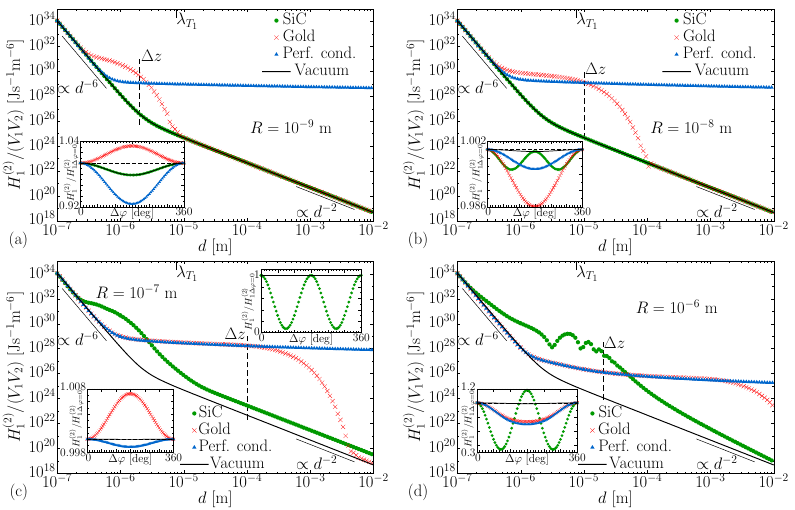}
\end{center}
\caption{\label{fig:HTpar}Heat transfer (normalized by particles' volumes) from SiC particle $ 1 $ at temperature $ T_1 = 300 \ \textrm{K} $ (the corresponding thermal wavelength $ \lambda_{T_1} = 7.63 \times 10^{-6} \ \textrm{m} $) to SiC particle $ 2 $ in the presence of a cylinder, as a function of interparticle distance $ d $. The particles are placed parallel to the cylinder axis at distance $ h = 10^{-7} \ \textrm{m} $ from the cylinder surface (see Fig.~\ref{fig:SystemHTpar}). The results are given for different radii $ R $ and materials of the cylinder, and they are compared to the case of the particles in vacuum. Inset graphs show the angular dependence of the HT, where one particle is rotated by an angle $ \Delta\varphi $ relative to the other, whereas radial and $z$ components remain unchanged (see Fig.~\ref{fig:SystemHTpar}); the result is normalized by the  case shown in the main graph ($ \Delta\varphi=0 $) and given for a specific distance along the $z$ axis, $\Delta z$, indicated in the main figures as vertical dashed lines; the point and color codes are the same as in the main figures, the horizontal dashed lines correspond to $ 1 $.}
\end{figure*}

For demonstration of the HT, we choose SiC particles, with permittivities given by Eq.~\eqref{eq:epsilon_SiC}; the temperature of particle $ 1 $ is $ T_1 = 300 \ \textrm{K} $. As in Sec.~\ref{sec:HR}, we consider three  materials of a cylinder: SiC, gold, and a perfect conductor.

\begin{figure}[!t]
\begin{center}
\includegraphics[width=1.0\linewidth]{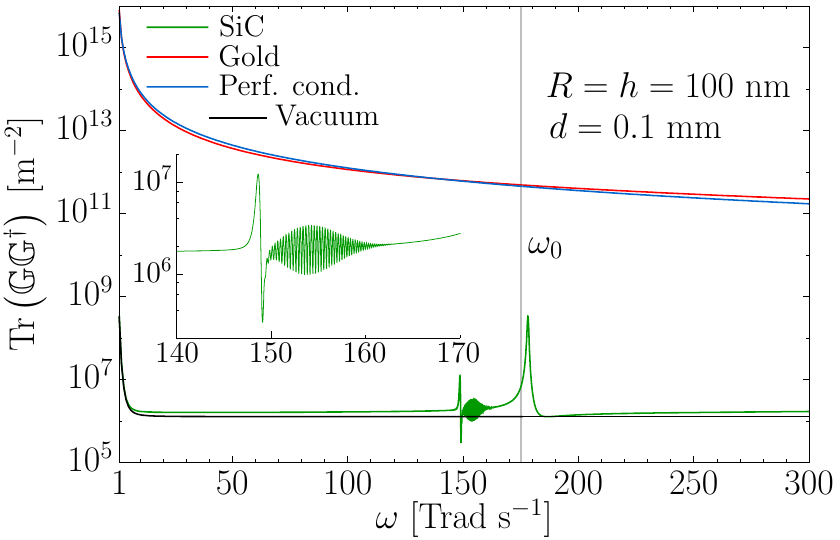}
\end{center}
\caption{\label{fig:SpecHT}$ \Tr\left[\mathbb{G}(\vct{r}_1, \vct{r}_2)\mathbb{G}^{\dagger}(\vct{r}_1, \vct{r}_2)\right] $ as a function of frequency, which is a part of the interparticle heat transfer spectrum related to the influence of a cylinder [see Eq.~\eqref{eq:HT}]. Here, we consider a parallel configuration (see Fig.~\ref{fig:SystemHTpar}) with $ R = h = 100 \ \textrm{nm} $ and $ d = 0.1 \ \textrm{mm} $, corresponding to the vertical dashed line in Fig.~\ref{fig:HTpar}(c). The vertical line shows the resonance frequency $ \omega_0 $ of the SiC particle polarizability, selecting a major contribution to the total heat transfer [see Eqs.~\eqref{eq:polarizability},~\eqref{eq:epsilon_SiC}, and~\eqref{eq:HT}]. The inset shows a zoomed version of the main plot for $ \omega \in [140, 170] \ \textrm{Trad} \ \textrm{s}^{-1} $.}
\end{figure}

In this section, we consider the configuration depicted in Fig.~\ref{fig:SystemHTpar}, where the interparticle distance line is either parallel (for azimuthal angle $\Delta\varphi=0$) or almost parallel ($\Delta\varphi \neq 0$, but $ \Delta z \gg 2r $, such that $ d \approx \Delta z $) to the cylinder axis. Both particles are placed at a distance $ h $ from the cylinder surface; we set $ h = 10^{-7} \ \textrm{m} $, i.e., $ h \ll \lambda_{T_1} $, in order to have a strong coupling between the particles and cylinder. The GF is given by Eqs.~\eqref{eq:G0Parallel} and~\eqref{eq:GTParallel} for $ \Delta\varphi = 0 $, and by Eqs.~\eqref{eq:G0CylindricalFinalEqualr} and~\eqref{eq:GTEqualr} for finite $ \Delta\varphi $. For $ \Delta\varphi = 0 $, this configuration was studied in Ref.~\citep{Asheichyk2022}. It was found that, for a perfectly conducting cylinder, the HT decays logarithmically with the interparticle distance $ d $ (for large $ d $), thus being many orders of magnitude larger than for isolated particles. This highly efficient energy transport is attributed to the system geometry: The cylinder acts as a waveguide, transferring the near-field energy in the preferred direction. Here, we study this configuration in more detail, including SiC and gold cylinders, as well as effects of finite angle $\Delta\varphi$.

The HT is given in Fig.~\ref{fig:HTpar} as a function of $ d $ for different radii and materials of the cylinder. Overall, the cylinder greatly enhances the HT. When $ R $ is small [$ R \ll h $ in Figs.~\ref{fig:HTpar}(a) and ~\ref{fig:HTpar}(b)], a SiC cylinder has little effect, whereas a gold cylinder yields a large enhancement, which appears at larger values of $d$ and vanishes at larger values of $d$, the larger $ R $ is. The vanishing occurs via an exponential decay to the vacuum result. Compared to gold, the HT with a perfectly conducting cylinder deviates from the vacuum HT at a larger near-field distance, meaning that gold wins in terms of efficiency in the near field. However, an exponential decay does not appear for a perfect conductor, thanks to the absence of material losses. Instead, the HT decays logarithmically for any large $ d $~\cite{Asheichyk2022}, surpassing gold, SiC, or vacuum cases by more than $ 10 $ orders of magnitude in the far field (for the range of $d$ shown).

When $ R = h = 10^{-7} \ \textrm{m} $ [Fig.~\ref{fig:HTpar}(c)], a SiC cylinder strongly affects the HT, being a few orders of magnitude better than gold and a perfect conductor for near-field $ d \gtrsim h $. For intermediate distances, the SiC curve goes below the gold and perfect conductor curves, but it is still above the vacuum result. When $ d $ is large compared to the thermal wavelength, the HT scales similar to the vacuum one, i.e., $ \propto d^{-2} $, but it is about seven times larger; the ultimate behavior for $ d \to \infty $ remains unknown. The HT in the presence of a gold cylinder features a plateau starting at $ d \approx 1 \ \mu\textrm{m} $ and ending by an exponential decay in the far field. This plateau is longer in $ d $ but smaller in the HT amplitude compared to that observed for smaller $ R $ [see Figs.~\ref{fig:HTpar}(a) and~\ref{fig:HTpar}(b)]. Until the exponential drop, it almost repeats the logarithmically decaying perfect conductor curve (gold has a slightly larger HT). The latter is also smaller in amplitude compared to the HT for smaller $ R $ (a detailed $ R $ dependence can be found in Ref.~\citep{Asheichyk2022}).

We analyze the case of $ R = h = 10^{-7} \ \textrm{m} $ and $ d = 10^{-4} \ \textrm{m} $ in more detail, plotting the spectrum of $ \Tr\left(\mathbb{G}\mathbb{G}^{\dagger}\right) $ in Fig.~\ref{fig:SpecHT}. The amplification for SiC observed in Fig.~\ref{fig:HTpar}(c) can be explained by the resonance of $ \Tr\left(\mathbb{G}\mathbb{G}^{\dagger}\right) $ close to $\omega_0$ (the resonance of $ \alpha_{\textrm{SiC}} $). The small mismatch of the respective resonance frequencies of particle and cylinder implies that the HT can be increased even further when using (for either particles or a cylinder) a material with slightly different optical properties. Similar arguments can be applied for the HR in the presence of a SiC cylinder (see the peak in Fig.~\ref{fig:SpecHR}). Interestingly, the spectrum also features a \enquote{pulse} at $ \omega \approx 149 \ \textrm{Trad} \ \textrm{s}^{-1} $ (where the real part of $ \varepsilon_{\textrm{SiC}} $ changes sign), followed by a \enquote{wave packet.} For gold and a perfect conductor, the spectrum is a smooth function of $ \omega $, with a much larger amplitude compared to vacuum or SiC, which explains the large amplification in Fig.~\ref{fig:HTpar}. This smooth behavior is different compared to the case of the particles inside a cylindrical gold cavity, where the spectrum features geometry-induced resonances~\cite{Asheichyk2023}.

For $ R = 10^{-6} \ \textrm{m} $ in Fig.~\ref{fig:HTpar}(d), i.e., when the radius becomes comparable to the skin depth of SiC, the HT in the presence of a SiC cylinder shows an oscillatory behavior for intermediate $ d $, being much larger than the HT in vacuum or with a gold or perfectly conducting cylinder present. The  oscillations, whose origin remains partly elusive, fade away with an increase of $ d $, and when $ d \gg \lambda_{T_1} $ the HT decays monotonically, approaching the vacuum result (we cannot exclude that it goes below the vacuum curve for $ d \gg 1 \ \textrm{cm} $, as was observed for a SiC plate~\cite{Messina2018}). Similar to Fig.~\ref{fig:HTpar}(c), gold and a perfect conductor show almost identical results until the exponential drop for gold. Compared to smaller $ R $, the logarithmic plateau spans a wider range of $ d $ but has a smaller amplitude. With a further increase of $ R $, the HT for all materials is expected to converge to the HT in the presence of a plate (the tendency can already be observed in Fig.~\ref{fig:HTpar} if one compares to the plate results~\cite{Dong2018, Messina2018, Asheichyk2018}).

How does the HT change if the symmetry of the configuration is violated, i.e., for finite azimuthal angle $\Delta\varphi$? For finite $\Delta\varphi$, the interparticle distance $ d $ is given by $ d = \sqrt{2r^2(1-\cos\Delta\varphi) + (\Delta z)^2} $, where $ r = R + h $ is the radial coordinate of each particle, and $ \Delta z $ is the separation along the cylinder axis, so that $d=\Delta z$ for $\Delta\varphi=0$ (see Fig.~\ref{fig:SystemHTpar}). To reveal the pure effects of  rotation, and to minimize  the effect of the distance change, we consider $ \Delta z \gg 2r $, such that $ d \approx \Delta z $, i.e., $ d $ changes insignificantly with $\Delta\varphi$.

The angular dependent HT, normalized by the HT at $\Delta\varphi=0$, $H_{1\Delta\varphi=0}^{(2)} $, is given in the inset plots of Fig.~\ref{fig:HTpar} as a function of $ \Delta\varphi $. The corresponding $ \Delta z $ is indicated via vertical dashed lines in the main plots, which also displays the reference value of the HT. We note that the results are symmetric with respect to $ \Delta\varphi = 180^{\circ} $, as expected.

In the inset of Fig.~\ref{fig:HTpar}(a), $ R = 10^{-9} \ \textrm{m} $ and $ \Delta z = 2 \times 10^{-6} \ \textrm{m} $. The angular dependence for SiC is the same as the vacuum dependence (i.e., rotating the particle with no cylinder present, such that the HT decreases only due to increase of $ d $), meaning that such a thin SiC cylinder is transparent. Rotating the particle around a gold or perfectly conducting cylinder leads to a significant, yet no more than $ 1.09 $-fold, change of the HT. Interestingly, the rotation around a gold cylinder enhances the HT (with a maximum reached at $ \Delta\varphi = 180^{\circ} $), whereas the HT is decreased when the particle is rotated around a perfectly conducting cylinder (where $ \Delta\varphi = 180^{\circ} $ gives the minimum).

For $ R = 10^{-8} \ \textrm{m} $ at $ \Delta z = 10^{-5} \ \textrm{m} $ [see the inset of Fig.~\ref{fig:HTpar}(b)], all the curves are below $ 1 $, i.e., the rotation decreases the HT for all considered materials. A SiC cylinder gives a minimum at $ \Delta\varphi = 90^{\circ} $ (and hence also symmetrically at $ \Delta\varphi = 270^{\circ} $) and a local maximum at $ \Delta\varphi = 180^{\circ} $ coinciding with the minimum of the vacuum result. Minima at $ \Delta\varphi = 180^{\circ} $ appear for gold and a perfect conductor; for gold, the minimum is more pronounced.

For $ R = 10^{-7} \ \textrm{m} $, we choose $ \Delta z = 10^{-4} \ \textrm{m} $. Here, the rotation around a SiC cylinder leads to a large change of the HT [see the right inset of Fig.~\ref{fig:HTpar}(c)]. The ratio at the minimum is $ H_1^{(2)}(\Delta\varphi = 90^{\circ})/H_{1\Delta\varphi=0}^{(2)} = 0.0674 $, i.e., rotating the particle by $ 90^{\circ} $ suppresses the HT by $ 15 $ times. In contrast, with gold or a perfect conductor, there is only a little effect [see the left inset of Fig.~\ref{fig:HTpar}(c)]: A maximum with $ H_1^{(2)}(\Delta\varphi = 180^{\circ})/H_{1\Delta\varphi=0}^{(2)} = 1.0072 $ appears for gold, and a minimum with $ H_1^{(2)}(\Delta\varphi = 180^{\circ})/H_{1\Delta\varphi=0}^{(2)} = 0.9988 $ is observed for a perfect conductor.

When $ R = 10^{-6} \ \textrm{m} $ and $ \Delta z = 2 \times 10^{-5} $, the angular dependence for SiC is even more nontrivial, as can be seen in the inset plot of Fig.~\ref{fig:HTpar}(d). A minimum with the approximate value of $ 0.335 $ is reached at $ \Delta\varphi \approx 85^{\circ} $. It is followed by a maximum at $ \Delta\varphi = 180^{\circ} $ with the value of $ 1.1805 $. Compared to the previously considered $ R $ and $ \Delta z $, gold and a perfect conductor have now stronger minima at $ \Delta\varphi = 180^{\circ} $: $ H_1^{(2)}(\Delta\varphi = 180^{\circ})/H_{1\Delta\varphi=0}^{(2)} = 0.7309 $ for gold and $ H_1^{(2)}(\Delta\varphi = 180^{\circ})/H_{1\Delta\varphi=0}^{(2)} = 0.6918 $ for the perfect conductor.

Overall, we can conclude that $ H_1^{(2)}/H_{1\Delta\varphi=0}^{(2)} \leq 1 $ for a perfectly conducting cylinder, with a minimum at $ \Delta\varphi = 180^{\circ} $, whereas gold shows either a minimum or maximum at this angle, such that the ratio is either smaller or larger than $ 1 $, respectively. For the parameters studied, no more than an $ 8 \% $ and a $ 31 \% $ change of the HT is observed for a metallic nano- and microwire, respectively. This  indicates that a well-conducting cylinder transports the energy mainly via surface modes with little angular dependence, i.e., the mode $n=0$. For a SiC cylinder, the angular dependence is stronger and, the larger $R$ is, the modes with larger $n$ contribute.

\section{Heat transfer: perpendicular configuration}
\label{sec:HT_perpendicular}
\begin{figure}[!b]
\begin{center}
\includegraphics[width=0.95\linewidth]{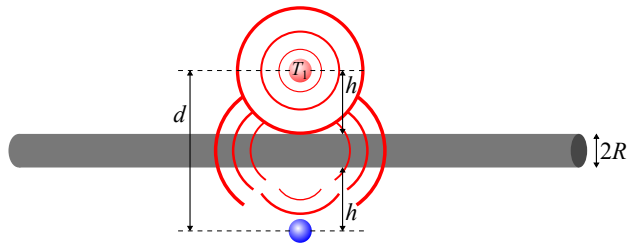}
\end{center}
\caption{\label{fig:SystemHTperp}Radiative heat transfer from particle $ 1 $ at temperature $ T_1 $ to particle $ 2 $, placed close to an infinitely long cylinder perpendicular to the cylinder axis. The cylinder either enhances or blocks the transfer, depending on its material (see Fig.~\ref{fig:HTperp}).}
\end{figure}

\begin{figure*}[!t]
\begin{center}
\includegraphics[width=0.8\linewidth]{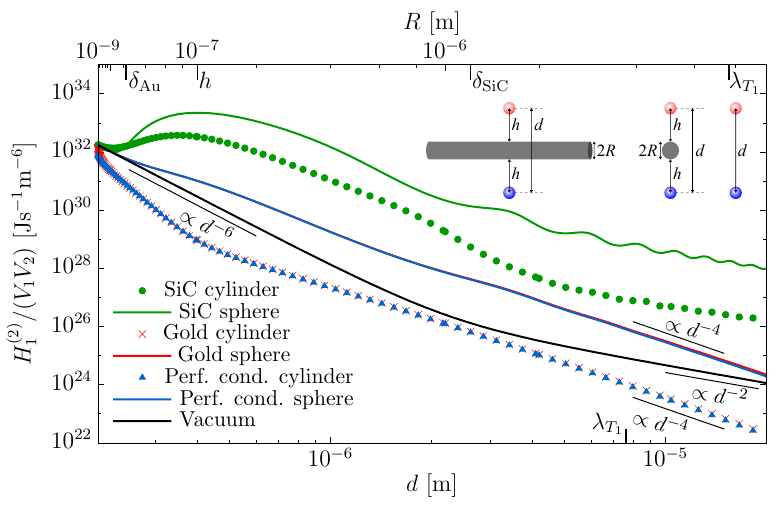}
\end{center}
\caption{\label{fig:HTperp}Heat transfer (normalized by particles' volumes) from SiC particle $ 1 $ at temperature $ T_1 = 300 \ \textrm{K} $ to SiC particle $ 2 $ placed perpendicular to a cylinder or on opposite sides of a sphere, or with no objects present (see the sketch), as a function of interparticle distance $ d $. The results are given for different materials of the cylinder and sphere. The distance from each particle to the cylinder (sphere) surface is $ h = 10^{-7} \ \textrm{m} $; the upper axis gives the corresponding radius of the object. On the lower axis, we show the thermal wavelength, while the upper axis also contains $ h $ and the skin depths of SiC and gold. The results for the sphere are taken from Ref.~\cite{Asheichyk2017}.}
\end{figure*}

Another important configuration is depicted in Fig.~\ref{fig:SystemHTperp}: the interparticle distance line is perpendicular to the cylinder axis (corresponding to $\Delta z=0$ and $ \Delta\varphi = 180^{\circ} $ in Fig.~\ref{fig:SystemHTpar}). Both particles are placed at a distance $ h $ from the cylinder surface, such that $ d = 2(R+h) $. The GF for this configuration is given by Eqs.~\eqref{eq:G0Perpendicular} and~\eqref{eq:GTPerpendicular}. Keeping small $ h =  10^{-7} \ \textrm{m} $ fixed, we aim to investigate how the HT depends on $ d $ when $ R $ is increased, as was done for the HT between the particles in the presence of a sphere in Ref.~\cite{Asheichyk2017} (see Fig.~5 there).

Figure~\ref{fig:HTperp} shows the HT as a function of $ d $ (lower horizontal axis) for different materials of the cylinder, compared to the HT in the presence of a sphere (with the same $ R $) and the HT for isolated particles; see the sketch. In the upper horizontal axis, we show the corresponding $ R $. 

The HT in the presence of a SiC cylinder is largely (by two to three orders of magnitude) above the vacuum HT for any considered $ R $ comparable to or larger than $ h $. The curve has a local maximum at $ R \approx 0.7h $, and it approaches the vacuum result when $ R \ll h $. A SiC sphere has a similar effect, but with a larger amplification, a local maximum at $ R \approx h $, and pronounced oscillations for $ R > \delta_{\textrm{SiC}} $~\cite{Asheichyk2017}.

On the contrary, a gold or perfectly conducting cylinder blocks the HT. Strongest suppression of the transfer appears at $ R \approx h $ (about $ 30 $ times) and for $ d > \lambda_{T_1} $ (here the HT scales approximately as $ \sim d^{-4} $ within the studied range of $ d $), whereas the effect is minimized at intermediate $ R \approx 1 \ \mu\textrm{m} $. Interestingly, even a very thin ($ R \ll h $) metallic cylinder strongly affects the HT. Here, the difference between gold and a perfect conductor becomes pronounced: when $ R \ll \delta_{\textrm{Au}} $, a gold cylinder becomes transparent, while a perfectly conducting cylinder suppresses the HT by more than twice even when $ R = 1 \ \textrm{nm} $. Putting a metallic sphere between the particles leads to a totally different behavior, from the HT being unaffected for small $ R $ to a large enhancement for $ R \in [h, \lambda_{T_1}] $ (when $ R \gg \lambda_{T_1} $, the HT goes below the vacuum result)~\cite{Asheichyk2017}.

The strong blocking by a cylinder may be understood from energy considerations. The cylinder is able to transport energy away to infinity, like a lightning rod. Such transport is impossible for the sphere, which thus is much worse in terms of blocking. 

It is worth noticing that the convergence of the HT with respect to the number of multipoles can be strongly nonmonotonic (for both parallel and perpendicular configurations), as was also observed for the HT in the presence of a sphere (see Fig.~6 in Ref.~\cite{Asheichyk2017}).

\section{Conclusion}
\label{sec:Conclusion}
In this paper, we studied heat radiation of a small particle and radiative heat transfer between small particles in the presence of a cylinder. Assumptions of small particle size and infinite extension of a cylinder along its axis allows us to reduce the problem to the study of the semi-analytical Green's function of a cylinder. Analyzing this Green's function in detail (as presented in the Appendix), we investigated several geometrical configurations, both symmetric (paradigmatic) and nonsymmetric, considering different materials for a cylinder.

A SiC particle placed close to a nanowire radiates much stronger than when being isolated, with the effect being strongest for a SiC nanowire. Notably, even a perfectly conducting cylinder, which absorbs no energy, strongly enhances HR of a closeby nanoparticle, which we attribute to excitation of waves traveling to infinity. The energy transfer between two particles along a metallic cylinder outperforms the transfer in vacuum by several orders of magnitude by virtue of slowly decaying surface waves. A SiC cylinder can be preferable for intermediate (micron-range) interparticle distances, showing a nonmonotonic dependence on the distance. These phenomena are stable in the sense that they vary little when the configuration is imperfect (particles are located at slightly different distances from the cylinder surface or one of them is slightly rotated around a cylinder), which is relevant for experiments, as it is difficult to achieve a perfect alignment in practice. Yet a large relative azimuthal angle can change the heat transfer dramatically in the case of SiC, but still mildly in the case of gold or a perfect conductor. This demonstrates that different wave modes are responsible for transfer for these materials. Placing a metallic cylinder between the particles blocks the heat transfer like a lightning rod, whereas a SiC cylinder acts as an amplifier.

The Green's function analyzed in the Appendix can be used to study other geometries and materials, including magnetic response of a cylinder. Future work may also consider electromagnetic conductivity of a cylinder, using a recent theory of electromagnetic heat transport in dissipative media~\cite{Kruger2024}, and consider the heat transfer between the particle and the cylinder.

\begin{acknowledgments}
We thank T.~Emig, D.~Gelbwaser-Klimovsky, N.~Graham, M.~Kardar, and R.~Messina for useful discussions. R.~Messina is also acknowledged for a critical reading of the manuscript. This work was funded by the Deutsche Forschungsgemeinschaft (DFG, German Research Foundation) through the Walter Benjamin Fellowship (Project No. 453458207).
\end{acknowledgments}

% MAIN_PART_end ------------------------------------------

% APPENDIX_begin ------------------------------------------

\appendix*

\section{Green's function of an infinitely long cylinder}
\label{app:sec:GF}
We work in a cylindrical coordinate system $ (r, \varphi, z) $ and consider an infinitely long cylinder of radius $ R $, whose symmetry axis coincides with the $ z $ axis (see Figs.~\ref{fig:SystemHR},~\ref{fig:SystemHTpar}, and~\ref{fig:SystemHTperp}). We aim to find the GF $ \mathbb{G}(\vct{r},\vct{r}') $ of the cylinder (more precisely, the electric GF, a part of the electromagnetic set of the GFs~\cite{Dong2017_1, Messina2018, Tai1994, Eckhardt1984}), where both radius vectors $ \vct{r} $ and $ \vct{r}' $ lie outside the cylinder.

\subsection{Free Green's function in cylindrical coordinates}
\label{app:subsec:G0cylindrical}
The GF of a cylinder contains the GF of free space $ \mathbb{G}_0 $ [see Eq.~\eqref{eq:G_separation}]. In a Cartesian coordinate system, this GF is well known in closed form~\cite{Dong2017_1, Messina2018, Tai1994, Asheichyk2017, Tsang2004, VanBladel1961, Yaghjian1980, Weiglhofer1989} ($ \mathbb{G}_0 $ and $ \widetilde{\mathbb{G}}_0 $ denote the free GF in a cylindrical and a Cartesian coordinate system, respectively):
\begin{align}
\notag \widetilde{\mathbb{G}}_0(\vct{r}, \vct{r}') = & -\frac{1}{3k^2}\mathcal{I}\delta^{(3)}(\vct{r}-\vct{r}')\\
\notag & + \frac{e^{ikd}}{4\pi k^2d^5}\Big[d^2(-1+ikd+k^2d^2)\mathcal{I}\\
& + (3-3ikd-k^2d^2)(\vct{r}-\vct{r}')\otimes(\vct{r}-\vct{r}')\Big],
\label{eq:G0Cartesian}
\end{align}
where $ d = |\vct{r}-\vct{r}'| $ is the distance between the points, $ k = \frac{\omega}{c} $ is the wave number (the amplitude of the wave vector), $ \mathcal{I} $ is the $ 3 \times 3 $ identity matrix, and the symbol $ \otimes $ denotes the dyadic product. Note the following properties for $ \widetilde{\mathbb{G}}_0 $:
\begin{subequations}
\begin{alignat}{1}
& \widetilde{\mathbb{G}}_0(\vct{r}, \vct{r}') = \widetilde{\mathbb{G}}^T_0(\vct{r}', \vct{r}),\label{eq:G0CartProp1}\\
& \widetilde{\mathbb{G}}_0(\vct{r}, \vct{r}') = \widetilde{\mathbb{G}}_0(\vct{r}-\vct{r}'),\label{eq:G0CartProp2}\\
& \widetilde{\mathbb{G}}_0(\vct{r}, \vct{r}') = \widetilde{\mathbb{G}}^T_0(\vct{r}, \vct{r}'),\label{eq:G0CartProp3}\\
& \widetilde{\mathbb{G}}_0(\vct{r}, \vct{r}') = \widetilde{\mathbb{G}}_0(\vct{r}', \vct{r}).\label{eq:G0CartProp4}
\end{alignat}
\end{subequations}
Property~\eqref{eq:G0CartProp1} is the reciprocity condition, which takes place for any GF~\cite{Tai1994, Eckhardt1984, Tsang2004}. The delta function term in Eq.~\eqref{eq:G0Cartesian}, which contributes to the field at the source region~\cite{Asheichyk2017, Tai1994, Tsang2004, VanBladel1961, Yaghjian1980, Weiglhofer1989, Nevels2004}, is neglected in other expressions of this work, because it does not contribute to the quantities of interest.

To obtain the free GF in cylindrical coordinates, $ \mathbb{G}_0 $, one can apply the corresponding transformation to the GF in Eq.~\eqref{eq:G0Cartesian}~\cite{Nikitin2013}:
\begin{equation}
\mathbb{G}_0(\vct{r}, \vct{r}') = \mathbb{G}_0(r, \varphi, z, r', \varphi', z') = \mathbb{U}(\varphi)\widetilde{\mathbb{G}}_0\mathbb{U}^{-1}(\varphi'),
\label{eq:G0Cylindrical}
\end{equation}
where the transformation of the coordinates ($ x=r\cos{\varphi} $, $ y=r\sin{\varphi} $, $ z=z $, $ x'=r'\cos{\varphi'} $, $ y'=r'\sin{\varphi'} $, $ z'=z' $) in $ \widetilde{\mathbb{G}}_0 $ is made;
\begin{equation}
\mathbb{U}(\varphi) = 
\begin{pmatrix}
\cos\varphi & \sin\varphi & 0\\
-\sin\varphi & \cos\varphi & 0\\
0 & 0 & 1
\end{pmatrix}
\label{eq:U}
\end{equation}
is the transformation matrix satisfying $ \mathbb{U}^{-1} = \mathbb{U}^T $~\cite{Nikitin2013}. We get ($ \Delta\varphi \equiv \varphi' - \varphi $ and $ \Delta z \equiv z' - z $)
\onecolumngrid
\begin{align}
\notag & \mathbb{G}_0(\vct{r}, \vct{r}') = \frac{e^{ikd}}{4\pi k^2d^5}\Bigg\{d^2(-1+ikd+k^2d^2)
\begin{pmatrix}
\cos\Delta\varphi & -\sin\Delta\varphi & 0\\
\sin\Delta\varphi & \cos\Delta\varphi & 0\\
0 & 0 & 1
\end{pmatrix}\\
& + (3-3ikd-k^2d^2)
\begin{pmatrix}
\left(r'\cos\Delta\varphi-r\right)\left(r'-r\cos\Delta\varphi\right) & \left(r'\cos\Delta\varphi-r\right)r\sin\Delta\varphi & \left(r'\cos\Delta\varphi-r\right)\Delta z\\
\left(r'-r\cos\Delta\varphi\right)r'\sin\Delta\varphi & rr'\sin^2\Delta\varphi & r'\sin\Delta\varphi\Delta z\\
\left(r'-r\cos\Delta\varphi\right)\Delta z & r\sin\Delta\varphi\Delta z & (\Delta z)^2
\end{pmatrix}
\Bigg\},
\label{eq:G0CylindricalFinal}
\end{align}
\twocolumngrid
satisfying
\begin{subequations}
\begin{alignat}{1}
& \mathbb{G}_0(\vct{r}, \vct{r}') = \mathbb{G}^T_0(\vct{r}', \vct{r}),\label{eq:G0CylProp1}\\
& \mathbb{G}_0(\vct{r}, \vct{r}') = \mathbb{G}_0(r,r',\varphi-\varphi',z-z').\label{eq:G0CylProp2}
\end{alignat}
\end{subequations}
The $ (11) $ term in the second matrix in Eq.~\eqref{eq:G0CylindricalFinal} can also be written as $ \left(r'\cos\Delta\varphi-r\right)\left(r'-r\cos\Delta\varphi\right) = \left(d^2-(\Delta z)^2\right)\cos\Delta\varphi - rr'\sin^2\Delta\varphi $. The distance $ d $ in terms of cylindrical coordinates is expressed as $ d = \sqrt{r^2 + r'^2 - 2rr'\cos\Delta\varphi + (\Delta z)^2} $.

Note that $ \mathbb{G}_0 $ can also be written as an expansion in cylindrical waves~\cite{Tai1994, Tsang2004, Golyk2012, Rahi2009, Kruger2012} [similar to $ \mathbb{G}_{\mathbb{T}} $ in Eq.~\eqref{eq:GT_expansion}]. However, this representation has restrictions on the positions ($ r \neq r' $ or $ z \neq z' $), which do not allow to study important configurations (as those considered in Secs.~\ref{sec:HT_parallel} and~\ref{sec:HT_perpendicular}). In contrast, Eq.~\eqref{eq:G0CylindricalFinal} is valid for arbitrary positions.

Transformation~\eqref{eq:G0Cylindrical} can be written for any GF, and its structure implies that our quantities of interest, $ \Tr\Im\mathbb{G}(\vct{r},\vct{r}) $ and $ \Tr\left(\mathbb{G}(\vct{r},\vct{r}')\mathbb{G}^{\dagger}(\vct{r},\vct{r}')\right) $, do not depend on coordinate system (given that $ \mathbb{U}^{-1} = \mathbb{U}^T $). This in turn means that HR and HT [see Eqs.~\eqref{eq:HR} and~\eqref{eq:HT}] do not depend on coordinate system, which can be also stated \textit{a priori} from physical grounds. For $ \mathbb{G}_0 $, the traces read as~\cite{Asheichyk2017}
\begin{equation}
\Tr\Im\mathbb{G}_0(\vct{r},\vct{r}) = \frac{k}{2\pi},
\label{eq:TrImG0}
\end{equation}
\begin{equation}
\Tr\left(\mathbb{G}_0\mathbb{G}^{\dagger}_0\right) = \frac{1}{8\pi^2d^2}\left[1+\frac{1}{k^2d^2}+\frac{3}{k^4d^4}\right],
\label{eq:TrG0G0Dagger}
\end{equation}
where $ d $ is kept finite in Eq.~\eqref{eq:TrG0G0Dagger}.

\subsection{Outgoing cylindrical waves}
\label{app:subsec:Waves}
As shown in Eq.~\eqref{eq:GT_expansion}, the scattering part of the GF can be written in terms of the outgoing cylindrical waves and the scattering matrix of a cylinder. The waves read as~\cite{Tsang2004, Golyk2012}
\begin{subequations}
\begin{alignat}{1}
& \vct{M}^{\textrm{out}}_{n,k_z}(\vct{r}) = \left[\frac{in}{qr}H_n(qr)\vct{e}_r-H'_n(qr)\vct{e}_{\varphi}\right]e^{ik_zz+in\varphi},\label{eq:Mwave}\\
\notag & \vct{N}^{\textrm{out}}_{n,k_z}(\vct{r}) = \frac{1}{k}\Big[ik_zH'_n(qr)\vct{e}_r-\frac{nk_z}{qr}H_n(qr)\vct{e}_{\varphi}\\
& \ \ \ \ \ \ \ \ \ \ \ \ \ \ \ + qH_n(qr)\vct{e}_z\Big]e^{ik_zz+in\varphi},\label{eq:Nwave}
\end{alignat}
\end{subequations}
where $ \vct{M}^{\textrm{out}}_{n,k_z} $ and $ \vct{N}^{\textrm{out}}_{n,k_z} $ correspond to magnetic and electric waves of multipole order $ n \in \mathbb{Z} $, respectively. $ \vct{e}_r $, $ \vct{e}_{\varphi} $, and $ \vct{e}_z $ are the unit vectors. $ k_z \in \mathbb{R} $ is the $ z $ component of the wave vector, while $ q = \sqrt{k^2-k_z^2} $. When $ |k_z| \leq k $, $ q $ is real and non-negative, whereas when $ |k_z| > k $, $ q $ is complex with a zero real part and a positive imaginary part. $ H_n $ is the Hankel function of the first kind of order $ n $, and $ H'_n(qr) \equiv \frac{dH_n(qr)}{d(qr)} $ is the corresponding derivative.

\subsection{Scattering matrix}
\label{app:subsec:Tmatrix}
In our case, we need the scattering matrix which relates the incident and scattered fields outside a cylinder. The outside scattering matrix of an infinitely long cylinder is known analytically. For a cylinder made of a homogeneous isotropic material with dielectric permittivity $ \varepsilon $ and magnetic permeability $ \mu $,  the elements of the scattering matrix read~\cite{Golyk2012, Bohren2004, Noruzifar2011}
\begin{subequations}
\begin{alignat}{1}
& T_{n,k_z}^{MM} = -\frac{J_n(qR)}{H_n(qR)}\frac{\Delta_1\Delta_4-K^2}{\Delta_1\Delta_2-K^2},\label{eq:TMM}\\
& T_{n,k_z}^{NN} = -\frac{J_n(qR)}{H_n(qR)}\frac{\Delta_2\Delta_3-K^2}{\Delta_1\Delta_2-K^2},\label{eq:TNN}\\
& T_{n,k_z}^{MN} = T_{n,k_z}^{NM} = \frac{2i}{\pi\sqrt{\varepsilon\mu}\left[qRH_n(qR)\right]^2}\frac{K}{\Delta_1\Delta_2-K^2}.\label{eq:TMN}
\end{alignat}
\end{subequations}
Here, $ J_n $ is the Bessel function of order $ n $,
\begin{subequations}
\begin{alignat}{1}
& \Delta_1 = \frac{J'_n(q_{\varepsilon}R)}{q_{\varepsilon}RJ_n(q_{\varepsilon}R)}-\frac{1}{\varepsilon}\frac{H'_n(qR)}{qRH_n(qR)},\label{eq:Delta1}\\
& \Delta_2 = \frac{J'_n(q_{\varepsilon}R)}{q_{\varepsilon}RJ_n(q_{\varepsilon}R)}-\frac{1}{\mu}\frac{H'_n(qR)}{qRH_n(qR)},\label{eq:Delta2}\\
& \Delta_3 = \frac{J'_n(q_{\varepsilon}R)}{q_{\varepsilon}RJ_n(q_{\varepsilon}R)}-\frac{1}{\varepsilon}\frac{J'_n(qR)}{qRJ_n(qR)},\label{eq:Delta3}\\
& \Delta_4 = \frac{J'_n(q_{\varepsilon}R)}{q_{\varepsilon}RJ_n(q_{\varepsilon}R)}-\frac{1}{\mu}\frac{J'_n(qR)}{qRJ_n(qR)},\label{eq:Delta4}
\end{alignat}
\end{subequations}
and
\begin{equation}
K = \frac{nk_z}{\sqrt{\varepsilon\mu}kR^2}\left(\frac{1}{q_{\varepsilon}^2}-\frac{1}{q^2}\right),
\label{eq:K}
\end{equation}
where $ q_{\varepsilon} = \sqrt{\varepsilon\mu k^2-k_z^2} $.

In the limit of perfect conductivity (or reflectivity), the scattering matrix simplifies to~\cite{Rahi2009, Asheichyk2022}
\begin{subequations}
\begin{alignat}{1}
& \lim_{|\varepsilon|\to\infty} T_{n,k_z}^{MM} = -\frac{J'_n(qR)}{H'_n(qR)},\label{eq:TMMmirror}\\
& \lim_{|\varepsilon|\to\infty} T_{n,k_z}^{NN} = -\frac{J_n(qR)}{H_n(qR)},\label{eq:TNNmirror}\\
& \lim_{|\varepsilon|\to\infty} T_{n,k_z}^{MN} = T_{n,k_z}^{NM} = 0.\label{eq:TMNmirror}
\end{alignat}
\end{subequations}

The scattering matrix can be generalized to the case of an anisotropic material~\cite{Golyk2012}.

\subsection{Cylindrical waves expansion}
\label{app:subsec:Expansion}
The GF can be written as a sum of the free GF $ \mathbb{G}_0 $ and the scattering part $ \mathbb{G}_{\mathbb{T}} $:
\begin{equation}
\mathbb{G} = \mathbb{G}_0 + \mathbb{G}_{\mathbb{T}} = \mathbb{G}_0 + \mathbb{G}_0\mathbb{T}\mathbb{G}_0,
\label{eq:G_separation}
\end{equation}
where $ \mathbb{T} $ is the scattering operator of the cylinder~\cite{Golyk2012}, and the operator multiplication is understood in $ \mathbb{G}_0\mathbb{T}\mathbb{G}_0 $~\cite{Bimonte2017, Rahi2009, Asheichyk2017, Kruger2012, Muller2017}. Detailed information about electromagnetic operators can be found in Refs.~\cite{Bimonte2017, Rahi2009, Asheichyk2017, Kruger2012, Muller2017}. 

Substituting the expansion of the free GF in cylindrical waves~\cite{Tai1994, Tsang2004, Golyk2012, Rahi2009, Kruger2012} into Eq.~\eqref{eq:G_separation}, and using the relation between the scattering operator and the scattering matrix~\cite{Golyk2012, Rahi2009, Kruger2012}, one obtains~\cite{Golyk2012} 
\begin{align}
\notag \mathbb{G}_{\mathbb{T}} = & \ \frac{i}{8\pi}\sum_{P,P'}\sum_{n=-\infty}^{\infty}(-1)^n\\
& \times \int_{-\infty}^{\infty}dk_z\vct{P}^{\textrm{out}}_{n,k_z}(\vct{r})\otimes\vct{P}^{'\textrm{out}}_{-n,-k_z}(\vct{r}')T_{n,k_z}^{PP'},
\label{eq:GT_expansion}
\end{align}
where $ P, P' = \{M, N\} $, the waves $ \vct{P}^{\textrm{out}}_{n,k_z} $ are given by Eqs.~\eqref{eq:Mwave} and~\eqref{eq:Nwave}, and the matrix elements $ T_{n,k_z}^{PP'} $ can be found in Eqs.~\eqref{eq:TMM}--\eqref{eq:TMN}. Like $ \mathbb{G}_0 $, $ \mathbb{G}_{\mathbb{T}} $ is a function of $ r $, $ r' $, $ \varphi-\varphi' $, $ z-z' $.

\onecolumngrid
\subsection{Green's function}
\label{app:subsec:GF}

\subsubsection{Arbitrary positions}
\label{app:subsubsec:GF_arbitrary}
The free GF is given in Eq.~\eqref{eq:G0CylindricalFinal}. Performing the tensor product in Eq.~\eqref{eq:GT_expansion} and using $ H_{-n}(qr) = (-1)^nH_n(qr) $, $ H'_{-n}(qr) = (-1)^nH'_n(qr) $, $ T_{-n,k_z}^{MM} = T_{n,k_z}^{MM} $, $ T_{-n,k_z}^{NN} = T_{n,k_z}^{NN} $, $ T_{-n,k_z}^{MN} = -T_{n,k_z}^{MN} $, $ T_{n,-k_z}^{MM} = T_{n,k_z}^{MM} $, $ T_{n,-k_z}^{NN} = T_{n,k_z}^{NN} $, $ T_{n,-k_z}^{MN} = -T_{n,k_z}^{MN} $, together with the symmetry of the summation and integration, the scattering part can be obtained:
\begin{equation}
\mathbb{G}_{\mathbb{T}} = 
\begin{pmatrix}
G_{\mathbb{T}11} & G_{\mathbb{T}12} & G_{\mathbb{T}13}\\
G_{\mathbb{T}21} & G_{\mathbb{T}22} & G_{\mathbb{T}23}\\
G_{\mathbb{T}31} & G_{\mathbb{T}32} & G_{\mathbb{T}33}
\end{pmatrix},
\label{eq:GT}
\end{equation}
with
\begin{subequations}
\begin{alignat}{1}
\notag & G_{\mathbb{T}11} = \frac{i}{4\pi}\int_0^{\infty}dk_z\frac{k_z^2}{k^2}H_1(qr)H_1(qr')T_{0,k_z}^{NN}\cos(k_z\Delta z)\\
\notag & \ \ \ \ \ \ \ \ \ \ +\frac{i}{2\pi}\sum_{n=1}^{\infty}\int_0^{\infty}dk_z\Bigg\{\frac{n^2}{(qr)(qr')}H_n(qr)H_n(qr')T_{n,k_z}^{MM} + \frac{nk_z}{k}\left[\frac{1}{qr}H_n(qr)H'_n(qr')+\frac{1}{qr'}H'_n(qr)H_n(qr')\right]T_{n,k_z}^{MN}\\
&  \ \ \ \ \ \ \ \ \ \ \ \ \ \ \ \ \ \ \ \ \ \ \ \ \ \ \ \ \ \ \ \ \ \ \ \ +\frac{k_z^2}{k^2}H'_n(qr)H'_n(qr')T_{n,k_z}^{NN}\Bigg\}\cos(n\Delta\varphi)\cos(k_z\Delta z),\label{eq:GT11}\\
\notag & G_{\mathbb{T}12} = -\frac{i}{2\pi}\sum_{n=1}^{\infty}\int_0^{\infty}dk_z\Bigg\{\frac{n}{qr}H_n(qr)H'_n(qr')T_{n,k_z}^{MM} + \frac{k_z}{k}\left[\frac{n^2}{(qr)(qr')}H_n(qr)H_n(qr')+H'_n(qr)H'_n(qr')\right]T_{n,k_z}^{MN}\\
&  \ \ \ \ \ \ \ \ \ \ \ \ \ \ \ \ \ \ \ \ \ \ \ \ \ \ \ \ \ \ \ \ \ \ \ \ +\frac{nk_z^2}{k^2qr'}H'_n(qr)H_n(qr')T_{n,k_z}^{NN}\Bigg\}\sin(n\Delta\varphi)\cos(k_z\Delta z),\label{eq:GT12}\\
\notag & G_{\mathbb{T}13} = -\frac{i}{4\pi}\int_0^{\infty}dk_z\frac{qk_z}{k^2}H_1(qr)H_0(qr')T_{0,k_z}^{NN}\sin(k_z\Delta z)\\
& \ \ \ \ \ \ \ \ \ \ +\frac{i}{2\pi}\sum_{n=1}^{\infty}\int_0^{\infty}dk_z\Bigg\{\frac{n}{kr}H_n(qr)H_n(qr')T_{n,k_z}^{MN} + \frac{qk_z}{k^2}H'_n(qr)H_n(qr')T_{n,k_z}^{NN}\Bigg\}\cos(n\Delta\varphi)\sin(k_z\Delta z),\label{eq:GT13}\\
\notag & G_{\mathbb{T}21} = \frac{i}{2\pi}\sum_{n=1}^{\infty}\int_0^{\infty}dk_z\Bigg\{\frac{n}{qr'}H'_n(qr)H_n(qr')T_{n,k_z}^{MM} + \frac{k_z}{k}\left[\frac{n^2}{(qr)(qr')}H_n(qr)H_n(qr')+H'_n(qr)H'_n(qr')\right]T_{n,k_z}^{MN}\\
&  \ \ \ \ \ \ \ \ \ \ \ \ \ \ \ \ \ \ \ \ \ \ \ \ \ \ \ \ \ \ \ \ \ \ \ \ +\frac{nk_z^2}{k^2qr}H_n(qr)H'_n(qr')T_{n,k_z}^{NN}\Bigg\}\sin(n\Delta\varphi)\cos(k_z\Delta z),\label{eq:GT21}\\
\notag & G_{\mathbb{T}22} = \frac{i}{4\pi}\int_0^{\infty}dk_zH_1(qr)H_1(qr')T_{0,k_z}^{MM}\cos(k_z\Delta z)\\
\notag & \ \ \ \ \ \ \ \ \ \ +\frac{i}{2\pi}\sum_{n=1}^{\infty}\int_0^{\infty}dk_z\Bigg\{H'_n(qr)H'_n(qr')T_{n,k_z}^{MM} + \frac{nk_z}{k}\left[\frac{1}{qr}H_n(qr)H'_n(qr')+\frac{1}{qr'}H'_n(qr)H_n(qr')\right]T_{n,k_z}^{MN}\\
&  \ \ \ \ \ \ \ \ \ \ \ \ \ \ \ \ \ \ \ \ \ \ \ \ \ \ \ \ \ \ \ \ \ \ \ \ +\frac{n^2k_z^2}{k^2(qr)(qr')}H_n(qr)H_n(qr')T_{n,k_z}^{NN}\Bigg\}\cos(n\Delta\varphi)\cos(k_z\Delta z),\label{eq:GT22}\\
& G_{\mathbb{T}23} = \frac{i}{2\pi}\sum_{n=1}^{\infty}\int_0^{\infty}dk_z\Bigg\{\frac{q}{k}H'_n(qr)H_n(qr')T_{n,k_z}^{MN} + \frac{nk_z}{k^2r}H_n(qr)H_n(qr')T_{n,k_z}^{NN}\Bigg\}\sin(n\Delta\varphi)\sin(k_z\Delta z),\label{eq:GT23}\\
\notag & G_{\mathbb{T}31} = \frac{i}{4\pi}\int_0^{\infty}dk_z\frac{qk_z}{k^2}H_0(qr)H_1(qr')T_{0,k_z}^{NN}\sin(k_z\Delta z)\\
& \ \ \ \ \ \ \ \ \ \ -\frac{i}{2\pi}\sum_{n=1}^{\infty}\int_0^{\infty}dk_z\Bigg\{\frac{n}{kr'}H_n(qr)H_n(qr')T_{n,k_z}^{MN} + \frac{qk_z}{k^2}H_n(qr)H'_n(qr')T_{n,k_z}^{NN}\Bigg\}\cos(n\Delta\varphi)\sin(k_z\Delta z),\label{eq:GT31}\\
& G_{\mathbb{T}32} = \frac{i}{2\pi}\sum_{n=1}^{\infty}\int_0^{\infty}dk_z\Bigg\{\frac{q}{k}H_n(qr)H'_n(qr')T_{n,k_z}^{MN} + \frac{nk_z}{k^2r'}H_n(qr)H_n(qr')T_{n,k_z}^{NN}\Bigg\}\sin(n\Delta\varphi)\sin(k_z\Delta z),\label{eq:GT32}\\
& G_{\mathbb{T}33} = \frac{i}{4\pi}\int_0^{\infty}dk_z\frac{q^2}{k^2}H_0(qr)H_0(qr')T_{0,k_z}^{NN}\cos(k_z\Delta z) + \frac{i}{2\pi}\sum_{n=1}^{\infty}\int_0^{\infty}dk_z\frac{q^2}{k^2}H_n(qr)H_n(qr')T_{n,k_z}^{NN}\cos(n\Delta\varphi)\cos(k_z\Delta z),\label{eq:GT33}
\end{alignat}
\end{subequations}
satisfying, as $ \mathbb{G}_0 $, properties~\eqref{eq:G0CylProp1} and~\eqref{eq:G0CylProp2}, such that the full GF, $ \mathbb{G} = \mathbb{G}_0 + \mathbb{G}_{\mathbb{T}} $, also satisfies those properties.

\subsubsection{Equal radial coordinates}
\label{app:subsubsec:GF_equalr}
In the case of equal radial coordinates ($ r = r' = R + h $), the GF contains six independent elements. For $ \mathbb{G}_0 $ in Eq.~\eqref{eq:G0CylindricalFinal}, we get
\begin{align}
\notag & \mathbb{G}_0 = \frac{e^{ikd}}{4\pi k^2d^5}\Bigg\{d^2(-1+ikd+k^2d^2)
\begin{pmatrix}
\cos\Delta\varphi & -\sin\Delta\varphi & 0\\
\sin\Delta\varphi & \cos\Delta\varphi & 0\\
0 & 0 & 1
\end{pmatrix}\\
& + (3-3ikd-k^2d^2)
\begin{pmatrix}
-r^2\left(1-\cos\Delta\varphi\right)^2 & -r^2\left(1-\cos\Delta\varphi\right)\sin\Delta\varphi & -r\left(1-\cos\Delta\varphi\right)\Delta z\\
r^2\left(1-\cos\Delta\varphi\right)\sin\Delta\varphi & r^2\sin^2\Delta\varphi & r\sin\Delta\varphi\Delta z\\
r\left(1-\cos\Delta\varphi\right)\Delta z & r\sin\Delta\varphi\Delta z & (\Delta z)^2
\end{pmatrix}
\Bigg\},
\label{eq:G0CylindricalFinalEqualr}
\end{align}
where $ d = \sqrt{2r^2(1-\cos\Delta\varphi) + (\Delta z)^2} $.

The scattering part is obtained by setting $ r' = r $ in Eqs.~\eqref{eq:GT11}--\eqref{eq:GT33}:
\begin{equation}
\mathbb{G}_{\mathbb{T}} = 
\begin{pmatrix}
G_{\mathbb{T}11} & G_{\mathbb{T}12} & G_{\mathbb{T}13}\\
-G_{\mathbb{T}12} & G_{\mathbb{T}22} & G_{\mathbb{T}23}\\
-G_{\mathbb{T}13} & G_{\mathbb{T}23} & G_{\mathbb{T}33}
\end{pmatrix},
\label{eq:GTEqualr}
\end{equation}
with
\begin{subequations}
\begin{alignat}{1}
\notag & G_{\mathbb{T}11} = \frac{i}{4\pi}\int_0^{\infty}dk_z\frac{k_z^2}{k^2}H^2_1(qr)T_{0,k_z}^{NN}\cos(k_z\Delta z)\\
& \ \ \ \ \ \ \ \ \ \ +\frac{i}{2\pi}\sum_{n=1}^{\infty}\int_0^{\infty}dk_z\Bigg\{\frac{n^2}{(qr)^2}H^2_n(qr)T_{n,k_z}^{MM} + 2\frac{nk_z}{kqr}H_n(qr)H'_n(qr)T_{n,k_z}^{MN} + \frac{k_z^2}{k^2}\left[H'_n(qr)\right]^2T_{n,k_z}^{NN}\Bigg\}\cos(n\Delta\varphi)\cos(k_z\Delta z),\label{eq:GT11Equalr}\\
\notag & G_{\mathbb{T}22} = \frac{i}{4\pi}\int_0^{\infty}dk_zH^2_1(qr)T_{0,k_z}^{MM}\cos(k_z\Delta z)\\
& \ \ \ \ \ \ \ \ \ \ +\frac{i}{2\pi}\sum_{n=1}^{\infty}\int_0^{\infty}dk_z\Bigg\{[H'_n(qr)]^2T_{n,k_z}^{MM} + 2\frac{nk_z}{kqr}H_n(qr)H'_n(qr)T_{n,k_z}^{MN} +\frac{n^2k_z^2}{k^2(qr)^2}H^2_n(qr)T_{n,k_z}^{NN}\Bigg\}\cos(n\Delta\varphi)\cos(k_z\Delta z),\label{eq:GT22Equalr}\\
& G_{\mathbb{T}33} = \frac{i}{4\pi}\int_0^{\infty}dk_z\frac{q^2}{k^2}H^2_0(qr)T_{0,k_z}^{NN}\cos(k_z\Delta z) + \frac{i}{2\pi}\sum_{n=1}^{\infty}\int_0^{\infty}dk_z\frac{q^2}{k^2}H^2_n(qr)T_{n,k_z}^{NN}\cos(n\Delta\varphi)\cos(k_z\Delta z),\label{eq:GT33Equalr}\\
\notag & G_{\mathbb{T}12} = -\frac{i}{2\pi}\sum_{n=1}^{\infty}\int_0^{\infty}dk_z\Bigg\{\frac{n}{qr}H_n(qr)H'_n(qr)T_{n,k_z}^{MM} + \frac{k_z}{k}\left[\frac{n^2}{(qr)^2}H^2_n(qr)+\left[H'_n(qr)\right]^2\right]T_{n,k_z}^{MN}\\
&  \ \ \ \ \ \ \ \ \ \ \ \ \ \ \ \ \ \ \ \ \ \ \ \ \ \ \ \ \ \ \ \ \ \ \ \ +\frac{nk_z^2}{k^2qr}H_n(qr)H'_n(qr)T_{n,k_z}^{NN}\Bigg\}\sin(n\Delta\varphi)\cos(k_z\Delta z),\label{eq:GT12Equalr}\\
\notag & G_{\mathbb{T}13} = -\frac{i}{4\pi}\int_0^{\infty}dk_z\frac{qk_z}{k^2}H_0(qr)H_1(qr)T_{0,k_z}^{NN}\sin(k_z\Delta z)\\
& \ \ \ \ \ \ \ \ \ \ +\frac{i}{2\pi}\sum_{n=1}^{\infty}\int_0^{\infty}dk_z\Bigg\{\frac{n}{kr}H^2_n(qr)T_{n,k_z}^{MN} + \frac{qk_z}{k^2}H_n(qr)H'_n(qr)T_{n,k_z}^{NN}\Bigg\}\cos(n\Delta\varphi)\sin(k_z\Delta z),\label{eq:GT13Equalr}\\
& G_{\mathbb{T}23} = \frac{i}{2\pi}\sum_{n=1}^{\infty}\int_0^{\infty}dk_z\Bigg\{\frac{q}{k}H_n(qr)H'_n(qr)T_{n,k_z}^{MN} + \frac{nk_z}{k^2r}H^2_n(qr)T_{n,k_z}^{NN}\Bigg\}\sin(n\Delta\varphi)\sin(k_z\Delta z).\label{eq:GT23Equalr}
\end{alignat}
\end{subequations}

For a perfectly conducting cylinder, the elements of $ \mathbb{G}_{\mathbb{T}} $ take a simpler form thanks to compact expressions for the scattering matrix in Eqs.~\eqref{eq:TMMmirror}--\eqref{eq:TMNmirror}:
\begin{subequations}
\begin{alignat}{1}
\notag & \lim_{|\varepsilon|\to\infty} G_{\mathbb{T}11} = -\frac{i}{4\pi}\int_0^{\infty}dk_z\frac{k_z^2}{k^2}\frac{H^2_1(qr)J_0(qR)}{H_0(qR)}\cos(k_z\Delta z)\\
& \ \ \ \ \ \ \ \ \ \ \ \ \ \ \ \ \ \ -\frac{i}{2\pi}\sum_{n=1}^{\infty}\int_0^{\infty}dk_z\Bigg\{\frac{n^2}{(qr)^2}\frac{H^2_n(qr)J'_n(qR)}{H'_n(qR)} + \frac{k_z^2}{k^2}\frac{\left[H'_n(qr)\right]^2J_n(qR)}{H_n(qR)}\Bigg\}\cos(n\Delta\varphi)\cos(k_z\Delta z),\label{eq:GT11EqualrMirror}\\
\notag & \lim_{|\varepsilon|\to\infty} G_{\mathbb{T}22} = -\frac{i}{4\pi}\int_0^{\infty}dk_z\frac{H^2_1(qr)J_1(qR)}{H_1(qR)}\cos(k_z\Delta z)\\
& \ \ \ \ \ \ \ \ \ \ \ \ \ \ \ \ \ \ -\frac{i}{2\pi}\sum_{n=1}^{\infty}\int_0^{\infty}dk_z\Bigg\{\frac{[H'_n(qr)]^2J'_n(qR)}{H'_n(qR)} +\frac{n^2k_z^2}{k^2(qr)^2}\frac{H^2_n(qr)J_n(qR)}{H_n(qR)}\Bigg\}\cos(n\Delta\varphi)\cos(k_z\Delta z),\label{eq:GT22EqualrMirror}\\
& \lim_{|\varepsilon|\to\infty} G_{\mathbb{T}33} = -\frac{i}{4\pi}\int_0^{\infty}dk_z\frac{q^2}{k^2}\frac{H^2_0(qr)J_0(qR)}{H_0(qR)}\cos(k_z\Delta z) -\frac{i}{2\pi}\sum_{n=1}^{\infty}\int_0^{\infty}dk_z\frac{q^2}{k^2}\frac{H^2_n(qr)J_n(qR)}{H_n(qR)}\cos(n\Delta\varphi)\cos(k_z\Delta z),\label{eq:GT33EqualrMirror}\\
& \lim_{|\varepsilon|\to\infty} G_{\mathbb{T}12} = \frac{i}{2\pi}\sum_{n=1}^{\infty}\int_0^{\infty}dk_z\Bigg\{\frac{n}{qr}\frac{H_n(qr)H'_n(qr)J'_n(qR)}{H'_n(qR)} +\frac{nk_z^2}{k^2qr}\frac{H_n(qr)H'_n(qr)J_n(qR)}{H_n(qR)}\Bigg\}\sin(n\Delta\varphi)\cos(k_z\Delta z),\label{eq:GT12EqualrMirror}\\
\notag & \lim_{|\varepsilon|\to\infty} G_{\mathbb{T}13} = \frac{i}{4\pi}\int_0^{\infty}dk_z\frac{qk_z}{k^2}\frac{H_0(qr)H_1(qr)J_0(qR)}{H_0(qR)}\sin(k_z\Delta z)\\
& \ \ \ \ \ \ \ \ \ \ \ \ \ \ \ \ \ \ -\frac{i}{2\pi}\sum_{n=1}^{\infty}\int_0^{\infty}dk_z\frac{qk_z}{k^2}\frac{H_n(qr)H'_n(qr)J_n(qR)}{H_n(qR)}\cos(n\Delta\varphi)\sin(k_z\Delta z),\label{eq:GT13EqualrMirror}\\
& \lim_{|\varepsilon|\to\infty} G_{\mathbb{T}23} = -\frac{i}{2\pi}\sum_{n=1}^{\infty}\int_0^{\infty}dk_z\frac{nk_z}{k^2r}\frac{H^2_n(qr)J_n(qR)}{H_n(qR)}\sin(n\Delta\varphi)\sin(k_z\Delta z).\label{eq:GT23EqualrMirror}
\end{alignat}
\end{subequations}

\subsubsection{Equal angular coordinates}
\label{app:subsubsec:GF_equalphi}
In the case of equal angular coordinates ($ \varphi = \varphi' $), the GF contains five independent elements. For $ \mathbb{G}_0 $ in Eq.~\eqref{eq:G0CylindricalFinal}, we get
\begin{equation}
\mathbb{G}_0(\vct{r}, \vct{r}') = \frac{e^{ikd}}{4\pi k^2d^5}\Bigg\{d^2(-1+ikd+k^2d^2)
\begin{pmatrix}
1 & 0 & 0\\
0 & 1 & 0\\
0 & 0 & 1
\end{pmatrix}
+ (3-3ikd-k^2d^2)
\begin{pmatrix}
\left(r'-r\right)^2 & 0 & \left(r'-r\right)\Delta z\\
0 & 0 & 0\\
\left(r'-r\right)\Delta z & 0 & (\Delta z)^2
\end{pmatrix}
\Bigg\},
\label{eq:G0CylindricalFinalEqualphi}
\end{equation}
where $ d = \sqrt{\left(r'-r\right)^2 + (\Delta z)^2} $.

The scattering part is obtained by setting $ \varphi' = \varphi $ in Eqs.~\eqref{eq:GT11}--\eqref{eq:GT33}:
\begin{equation}
\mathbb{G}_{\mathbb{T}} = 
\begin{pmatrix}
G_{\mathbb{T}11} & 0 & G_{\mathbb{T}13}\\
0 & G_{\mathbb{T}22} & 0\\
G_{\mathbb{T}31} & 0 & G_{\mathbb{T}33}
\end{pmatrix},
\label{eq:GTEqualphi}
\end{equation}
with
\begin{subequations}
\begin{alignat}{1}
\notag & G_{\mathbb{T}11} = \frac{i}{4\pi}\int_0^{\infty}dk_z\frac{k_z^2}{k^2}H_1(qr)H_1(qr')T_{0,k_z}^{NN}\cos(k_z\Delta z)\\
\notag & \ \ \ \ \ \ \ \ \ \ +\frac{i}{2\pi}\sum_{n=1}^{\infty}\int_0^{\infty}dk_z\Bigg\{\frac{n^2}{(qr)(qr')}H_n(qr)H_n(qr')T_{n,k_z}^{MM} + \frac{nk_z}{k}\left[\frac{1}{qr}H_n(qr)H'_n(qr')+\frac{1}{qr'}H'_n(qr)H_n(qr')\right]T_{n,k_z}^{MN}\\
&  \ \ \ \ \ \ \ \ \ \ \ \ \ \ \ \ \ \ \ \ \ \ \ \ \ \ \ \ \ \ \ \ \ \ \ \ +\frac{k_z^2}{k^2}H'_n(qr)H'_n(qr')T_{n,k_z}^{NN}\Bigg\}\cos(k_z\Delta z),\label{eq:GT11Equalphi}\\
\notag & G_{\mathbb{T}22} = \frac{i}{4\pi}\int_0^{\infty}dk_zH_1(qr)H_1(qr')T_{0,k_z}^{MM}\cos(k_z\Delta z)\\
\notag & \ \ \ \ \ \ \ \ \ \ +\frac{i}{2\pi}\sum_{n=1}^{\infty}\int_0^{\infty}dk_z\Bigg\{H'_n(qr)H'_n(qr')T_{n,k_z}^{MM} + \frac{nk_z}{k}\left[\frac{1}{qr}H_n(qr)H'_n(qr')+\frac{1}{qr'}H'_n(qr)H_n(qr')\right]T_{n,k_z}^{MN}\\
&  \ \ \ \ \ \ \ \ \ \ \ \ \ \ \ \ \ \ \ \ \ \ \ \ \ \ \ \ \ \ \ \ \ \ \ \ +\frac{n^2k_z^2}{k^2(qr)(qr')}H_n(qr)H_n(qr')T_{n,k_z}^{NN}\Bigg\}\cos(k_z\Delta z),\label{eq:GT22Equalphi}\\
& G_{\mathbb{T}33} = \frac{i}{4\pi}\int_0^{\infty}dk_z\frac{q^2}{k^2}H_0(qr)H_0(qr')T_{0,k_z}^{NN}\cos(k_z\Delta z) + \frac{i}{2\pi}\sum_{n=1}^{\infty}\int_0^{\infty}dk_z\frac{q^2}{k^2}H_n(qr)H_n(qr')T_{n,k_z}^{NN}\cos(k_z\Delta z),\label{eq:GT33Equalphi}\\
\notag & G_{\mathbb{T}13} = -\frac{i}{4\pi}\int_0^{\infty}dk_z\frac{qk_z}{k^2}H_1(qr)H_0(qr')T_{0,k_z}^{NN}\sin(k_z\Delta z)\\
& \ \ \ \ \ \ \ \ \ \ +\frac{i}{2\pi}\sum_{n=1}^{\infty}\int_0^{\infty}dk_z\Bigg\{\frac{n}{kr}H_n(qr)H_n(qr')T_{n,k_z}^{MN} + \frac{qk_z}{k^2}H'_n(qr)H_n(qr')T_{n,k_z}^{NN}\Bigg\}\sin(k_z\Delta z),\label{eq:GT13Equalphi}\\
\notag & G_{\mathbb{T}31} = \frac{i}{4\pi}\int_0^{\infty}dk_z\frac{qk_z}{k^2}H_0(qr)H_1(qr')T_{0,k_z}^{NN}\sin(k_z\Delta z)\\
& \ \ \ \ \ \ \ \ \ \ -\frac{i}{2\pi}\sum_{n=1}^{\infty}\int_0^{\infty}dk_z\Bigg\{\frac{n}{kr'}H_n(qr)H_n(qr')T_{n,k_z}^{MN} + \frac{qk_z}{k^2}H_n(qr)H'_n(qr')T_{n,k_z}^{NN}\Bigg\}\sin(k_z\Delta z).\label{eq:GT31Equalphi}
\end{alignat}
\end{subequations}
For a perfectly conducting cylinder, we get
\begin{subequations}
\begin{alignat}{1}
\notag & G_{\mathbb{T}11} = -\frac{i}{4\pi}\int_0^{\infty}dk_z\frac{k_z^2}{k^2}\frac{H_1(qr)H_1(qr')J_0(qR)}{H_0(qR)}\cos(k_z\Delta z)\\
& \ \ \ \ \ \ \ \ \ \ -\frac{i}{2\pi}\sum_{n=1}^{\infty}\int_0^{\infty}dk_z\Bigg\{\frac{n^2}{(qr)(qr')}\frac{H_n(qr)H_n(qr')J'_n(qR)}{H'_n(qR)} +\frac{k_z^2}{k^2}\frac{H'_n(qr)H'_n(qr')J_n(qR)}{H_n(qR)}\Bigg\}\cos(k_z\Delta z),\label{eq:GT11EqualphiMirror}\\
\notag & G_{\mathbb{T}22} = -\frac{i}{4\pi}\int_0^{\infty}dk_z\frac{H_1(qr)H_1(qr')J_1(qR)}{H_1(qR)}\cos(k_z\Delta z)\\
& \ \ \ \ \ \ \ \ \ \ -\frac{i}{2\pi}\sum_{n=1}^{\infty}\int_0^{\infty}dk_z\Bigg\{\frac{H'_n(qr)H'_n(qr')J'_n(qR)}{H_n'(qR)} +\frac{n^2k_z^2}{k^2(qr)(qr')}\frac{H_n(qr)H_n(qr')J_n(qR)}{H_n(qR)}\Bigg\}\cos(k_z\Delta z),\label{eq:GT22EqualphiMirror}\\
& G_{\mathbb{T}33} = -\frac{i}{4\pi}\int_0^{\infty}dk_z\frac{q^2}{k^2}\frac{H_0(qr)H_0(qr')J_0(qR)}{H_0(qR)}\cos(k_z\Delta z) - \frac{i}{2\pi}\sum_{n=1}^{\infty}\int_0^{\infty}dk_z\frac{q^2}{k^2}\frac{H_n(qr)H_n(qr')J_n(qR)}{H_n(qR)}\cos(k_z\Delta z),\label{eq:GT33EqualphiMirror}\\
& G_{\mathbb{T}13} = \frac{i}{4\pi}\int_0^{\infty}dk_z\frac{qk_z}{k^2}\frac{H_1(qr)H_0(qr')J_0(qR)}{H_0(qR)}\sin(k_z\Delta z) -\frac{i}{2\pi}\sum_{n=1}^{\infty}\int_0^{\infty}dk_z\frac{qk_z}{k^2}\frac{H'_n(qr)H_n(qr')J_n(qR)}{H_n(qR)}\sin(k_z\Delta z),\label{eq:GT13EqualphiMirror}\\
& G_{\mathbb{T}31} = -\frac{i}{4\pi}\int_0^{\infty}dk_z\frac{qk_z}{k^2}\frac{H_0(qr)H_1(qr')J_0(qR)}{H_0(qR)}\sin(k_z\Delta z) +\frac{i}{2\pi}\sum_{n=1}^{\infty}\int_0^{\infty}dk_z\frac{qk_z}{k^2}\frac{H_n(qr)H'_n(qr')J_n(qR)}{H_n(qR)}\sin(k_z\Delta z).\label{eq:GT31EqualphiMirror}
\end{alignat}
\end{subequations}

\subsection{Green's function for parallel configuration}
\label{app:subsec:Parallel}
For configurations of interest, $ \mathbb{G}_0 $ and $ \mathbb{G}_{\mathbb{T}} $ can be simplified. For a parallel configuration (see Fig.~\ref{fig:SystemHTpar}), $ r = r' = R + h $, $ \varphi = \varphi' $ (i.e., $ \Delta\varphi = 0 $), and $ d = |z-z'| = |\Delta z| $. Giving this, $ \mathbb{G}_0 $ in Eq.~\eqref{eq:G0CylindricalFinalEqualr} becomes diagonal (with $ G_{011} = G_{022} $) and depends only on $ d $:
\begin{equation}
\mathbb{G}_0 = \frac{e^{ikd}}{4\pi k^2d^3}
\begin{pmatrix}
-1+ikd+k^2d^2 & 0 & 0\\
0 & -1+ikd+k^2d^2 & 0\\
0 & 0 & 2-2ikd
\end{pmatrix}.
\label{eq:G0Parallel}
\end{equation}

Without loss of generality, we consider $ z' > z $, such that $ \Delta z \equiv z' - z = d $. Using Eqs.~\eqref{eq:GT11Equalr}--\eqref{eq:GT23Equalr}, the scattering part for a parallel configuration can be obtained~\cite{Asheichyk2022}:
\begin{equation}
\mathbb{G}_{\mathbb{T}} = 
\begin{pmatrix}
G_{\mathbb{T}11} & 0 & G_{\mathbb{T}13}\\
0 & G_{\mathbb{T}22} & 0\\
-G_{\mathbb{T}13} & 0 & G_{\mathbb{T}33}
\end{pmatrix},
\label{eq:GTParallel}
\end{equation}
with
\begin{subequations}
\begin{alignat}{1}
\notag & G_{\mathbb{T}11} = \frac{i}{4\pi}\int_0^{\infty}dk_z\frac{k_z^2}{k^2}H^2_1(qr)T_{0,k_z}^{NN}\cos(k_zd)\\
& \ \ \ \ \ \ \ \ \ \ +\frac{i}{2\pi}\sum_{n=1}^{\infty}\int_0^{\infty}dk_z\Bigg\{\frac{n^2}{(qr)^2}H^2_n(qr)T_{n,k_z}^{MM} + 2\frac{nk_z}{kqr}H_n(qr)H'_n(qr)T_{n,k_z}^{MN} + \frac{k_z^2}{k^2}\left[H'_n(qr)\right]^2T_{n,k_z}^{NN}\Bigg\}\cos(k_zd),\label{eq:GT11Parallel}\\
\notag & G_{\mathbb{T}22} = \frac{i}{4\pi}\int_0^{\infty}dk_zH^2_1(qr)T_{0,k_z}^{MM}\cos(k_zd)\\
& \ \ \ \ \ \ \ \ \ \ +\frac{i}{2\pi}\sum_{n=1}^{\infty}\int_0^{\infty}dk_z\Bigg\{[H'_n(qr)]^2T_{n,k_z}^{MM} + 2\frac{nk_z}{kqr}H_n(qr)H'_n(qr)T_{n,k_z}^{MN} +\frac{n^2k_z^2}{k^2(qr)^2}H^2_n(qr)T_{n,k_z}^{NN}\Bigg\}\cos(k_zd),\label{eq:GT22Parallel}\\
& G_{\mathbb{T}33} = \frac{i}{4\pi}\int_0^{\infty}dk_z\frac{q^2}{k^2}H^2_0(qr)T_{0,k_z}^{NN}\cos(k_zd) + \frac{i}{2\pi}\sum_{n=1}^{\infty}\int_0^{\infty}dk_z\frac{q^2}{k^2}H^2_n(qr)T_{n,k_z}^{NN}\cos(k_zd),\label{eq:GT33Parallel}\\
\notag & G_{\mathbb{T}13} = -\frac{i}{4\pi}\int_0^{\infty}dk_z\frac{qk_z}{k^2}H_0(qr)H_1(qr)T_{0,k_z}^{NN}\sin(k_zd)\\
& \ \ \ \ \ \ \ \ \ \ +\frac{i}{2\pi}\sum_{n=1}^{\infty}\int_0^{\infty}dk_z\Bigg\{\frac{n}{kr}H^2_n(qr)T_{n,k_z}^{MN} + \frac{qk_z}{k^2}H_n(qr)H'_n(qr)T_{n,k_z}^{NN}\Bigg\}\sin(k_zd).\label{eq:GT13Parallel}
\end{alignat}
\end{subequations}
Note that interchanging the points $ \vct{r} $ and $ \vct{r}' $ is equivalent to replacing $ d $ with $ -d $ in Eqs.~\eqref{eq:GT11Parallel}--\eqref{eq:GT13Parallel}, which in turn is equivalent to making the transposition of $ \mathbb{G} $, in agreement with the reciprocity principle.

For a perfectly conducting cylinder, we get~\cite{Asheichyk2022}
\begin{subequations}
\begin{alignat}{1}
\notag & \lim_{|\varepsilon|\to\infty} G_{\mathbb{T}11} = -\frac{i}{4\pi}\int_0^{\infty}dk_z\frac{k_z^2}{k^2}\frac{H^2_1(qr)J_0(qR)}{H_0(qR)}\cos(k_zd)\\
& \ \ \ \ \ \ \ \ \ \ \ \ \ \ \ \ \ \ -\frac{i}{2\pi}\sum_{n=1}^{\infty}\int_0^{\infty}dk_z\Bigg\{\frac{n^2}{(qr)^2}\frac{H^2_n(qr)J'_n(qR)}{H'_n(qR)} + \frac{k_z^2}{k^2}\frac{\left[H'_n(qr)\right]^2J_n(qR)}{H_n(qR)}\Bigg\}\cos(k_zd),\label{eq:GT11ParallelMirror}\\
\notag & \lim_{|\varepsilon|\to\infty} G_{\mathbb{T}22} = -\frac{i}{4\pi}\int_0^{\infty}dk_z\frac{H^2_1(qr)J_1(qR)}{H_1(qR)}\cos(k_zd)\\
& \ \ \ \ \ \ \ \ \ \ \ \ \ \ \ \ \ \ -\frac{i}{2\pi}\sum_{n=1}^{\infty}\int_0^{\infty}dk_z\Bigg\{\frac{[H'_n(qr)]^2J'_n(qR)}{H'_n(qR)} +\frac{n^2k_z^2}{k^2(qr)^2}\frac{H^2_n(qr)J_n(qR)}{H_n(qR)}\Bigg\}\cos(k_zd),\label{eq:GT22ParallelMirror}\\
& \lim_{|\varepsilon|\to\infty} G_{\mathbb{T}33} = -\frac{i}{4\pi}\int_0^{\infty}dk_z\frac{q^2}{k^2}\frac{H^2_0(qr)J_0(qR)}{H_0(qR)}\cos(k_zd) -\frac{i}{2\pi}\sum_{n=1}^{\infty}\int_0^{\infty}dk_z\frac{q^2}{k^2}\frac{H^2_n(qr)J_n(qR)}{H_n(qR)}\cos(k_zd),\label{eq:GT33ParallelMirror}\\
& \lim_{|\varepsilon|\to\infty} G_{\mathbb{T}13} = \frac{i}{4\pi}\int_0^{\infty}dk_z\frac{qk_z}{k^2}\frac{H_0(qr)H_1(qr)J_0(qR)}{H_0(qR)}\sin(k_zd) -\frac{i}{2\pi}\sum_{n=1}^{\infty}\int_0^{\infty}dk_z\frac{qk_z}{k^2}\frac{H_n(qr)H'_n(qr)J_n(qR)}{H_n(qR)}\sin(k_zd).\label{eq:GT13ParallelMirror}
\end{alignat}
\end{subequations}

\subsection{Green's function for perpendicular configuration}
\label{app:subsec:Perpendicular}
For a perpendicular configuration (see Fig.~\ref{fig:SystemHTperp}), $ r = r' = R + h $, $ |\varphi - \varphi'| = \pi $, and $ z = z' $, with $ d = 2r = 2(R+h) $. Giving this, $ \mathbb{G}_0 $ in Eq.~\eqref{eq:G0CylindricalFinalEqualr} becomes diagonal (with $ G_{022} = -G_{033} $) and depends only on $ d $:
\begin{equation}
\mathbb{G}_0 = \frac{e^{ikd}}{4\pi k^2d^3}
\begin{pmatrix}
-2+2ikd & 0 & 0\\
0 & 1-ikd-k^2d^2 & 0\\
0 & 0 & -1+ikd+k^2d^2
\end{pmatrix}.
\label{eq:G0Perpendicular}
\end{equation}

The scattering part is also diagonal,
\begin{equation}
\mathbb{G}_{\mathbb{T}} = 
\begin{pmatrix}
G_{\mathbb{T}11} & 0 & 0\\
0 & G_{\mathbb{T}22} & 0\\
0 & 0 & G_{\mathbb{T}33}
\end{pmatrix},
\label{eq:GTPerpendicular}
\end{equation}
with
\begin{subequations}
\begin{alignat}{1}
\notag & G_{\mathbb{T}11} = \frac{i}{4\pi}\int_0^{\infty}dk_z\frac{k_z^2}{k^2}H^2_1(qr)T_{0,k_z}^{NN}\\
& \ \ \ \ \ \ \ \ \ \ +\frac{i}{2\pi}\sum_{n=1}^{\infty}\int_0^{\infty}dk_z\Bigg\{\frac{n^2}{(qr)^2}H^2_n(qr)T_{n,k_z}^{MM} + 2\frac{nk_z}{kqr}H_n(qr)H'_n(qr)T_{n,k_z}^{MN} + \frac{k_z^2}{k^2}\left[H'_n(qr)\right]^2T_{n,k_z}^{NN}\Bigg\}(-1)^n,\label{eq:GT11Perpendicular}\\
\notag & G_{\mathbb{T}22} = \frac{i}{4\pi}\int_0^{\infty}dk_zH^2_1(qr)T_{0,k_z}^{MM}\\
& \ \ \ \ \ \ \ \ \ \ +\frac{i}{2\pi}\sum_{n=1}^{\infty}\int_0^{\infty}dk_z\Bigg\{[H'_n(qr)]^2T_{n,k_z}^{MM} + 2\frac{nk_z}{kqr}H_n(qr)H'_n(qr)T_{n,k_z}^{MN} +\frac{n^2k_z^2}{k^2(qr)^2}H^2_n(qr)T_{n,k_z}^{NN}\Bigg\}(-1)^n,\label{eq:GT22Perpendicular}\\
& G_{\mathbb{T}33} = \frac{i}{4\pi}\int_0^{\infty}dk_z\frac{q^2}{k^2}H^2_0(qr)T_{0,k_z}^{NN} + \frac{i}{2\pi}\sum_{n=1}^{\infty}\int_0^{\infty}dk_z\frac{q^2}{k^2}H^2_n(qr)T_{n,k_z}^{NN}(-1)^n,\label{eq:GT33Perpendicular}
\end{alignat}
\end{subequations}
which differ from Eqs.~\eqref{eq:GT11Parallel}--\eqref{eq:GT33Parallel} by containing $ (-1)^n $ instead of $ \cos(k_zd) $. Interchanging the points $ \vct{r} $ and $ \vct{r}' $ does not affect $ \mathbb{G} $, and the transposition also has no effect due to diagonality of $ \mathbb{G} $, in agreement with the reciprocity principle.

For a perfectly conducting cylinder, we get
\begin{subequations}
\begin{alignat}{1}
\notag & \lim_{|\varepsilon|\to\infty} G_{\mathbb{T}11} = -\frac{i}{4\pi}\int_0^{\infty}dk_z\frac{k_z^2}{k^2}\frac{H^2_1(qr)J_0(qR)}{H_0(qR)}\\
& \ \ \ \ \ \ \ \ \ \ \ \ \ \ \ \ \ \ -\frac{i}{2\pi}\sum_{n=1}^{\infty}\int_0^{\infty}dk_z\Bigg\{\frac{n^2}{(qr)^2}\frac{H^2_n(qr)J'_n(qR)}{H'_n(qR)} + \frac{k_z^2}{k^2}\frac{\left[H'_n(qr)\right]^2J_n(qR)}{H_n(qR)}\Bigg\}(-1)^n,\label{eq:GT11PerpendicularMirror}\\
\notag & \lim_{|\varepsilon|\to\infty} G_{\mathbb{T}22} = -\frac{i}{4\pi}\int_0^{\infty}dk_z\frac{H^2_1(qr)J_1(qR)}{H_1(qR)}\\
& \ \ \ \ \ \ \ \ \ \ \ \ \ \ \ \ \ \ -\frac{i}{2\pi}\sum_{n=1}^{\infty}\int_0^{\infty}dk_z\Bigg\{\frac{[H'_n(qr)]^2J'_n(qR)}{H'_n(qR)} +\frac{n^2k_z^2}{k^2(qr)^2}\frac{H^2_n(qr)J_n(qR)}{H_n(qR)}\Bigg\}(-1)^n,\label{eq:GT22PerpendicularMirror}\\
& \lim_{|\varepsilon|\to\infty} G_{\mathbb{T}33} = -\frac{i}{4\pi}\int_0^{\infty}dk_z\frac{q^2}{k^2}\frac{H^2_0(qr)J_0(qR)}{H_0(qR)} -\frac{i}{2\pi}\sum_{n=1}^{\infty}\int_0^{\infty}dk_z\frac{q^2}{k^2}\frac{H^2_n(qr)J_n(qR)}{H_n(qR)}(-1)^n.\label{eq:GT33PerpendicularMirror}
\end{alignat}
\end{subequations}

\subsection{The trace of the imaginary part of the Green's function evaluated at equal points}
\label{app:subsec:TrImG}
To compute HR in Eq.~\eqref{eq:HR}, one has to know $ \Tr\Im\mathbb{G}(\vct{r},\vct{r}) $ (for a particle located at $ \vct{r}_1 $, $ \vct{r} = \vct{r}_1 $), which can be identified with the electric part of the local electromagnetic density of states at point $ \vct{r} $~\cite{Joulain2003, Joulain2005}. Note that $ \Tr\Im\mathbb{G}(\vct{r},\vct{r}) = \Tr\Im\mathbb{G}_0(\vct{r},\vct{r}) + \Tr\Im\mathbb{G}_{\mathbb{T}}(\vct{r},\vct{r}) $, where $ \Tr\Im\mathbb{G}_0(\vct{r},\vct{r}) $ is given by Eq.~\eqref{eq:TrImG0}. We hence need to concentrate only on $ \Tr\Im\mathbb{G}_{\mathbb{T}}(\vct{r},\vct{r}) $. Since $ \Tr\Im\mathbb{G}_{\mathbb{T}}(\vct{r},\vct{r}) = \Im\Tr\mathbb{G}_{\mathbb{T}}(\vct{r},\vct{r}) $, one can first evaluate $ \Tr\mathbb{G}_{\mathbb{T}}(\vct{r},\vct{r}) $. It can be found by inserting waves~\eqref{eq:Mwave} and~\eqref{eq:Nwave} into the trace of Eq.~\eqref{eq:GT_expansion},
\begin{equation}
\Tr\mathbb{G}_{\mathbb{T}}(\vct{r},\vct{r}) = \frac{i}{8\pi}\sum_{P,P'}\sum_{n=-\infty}^{\infty}(-1)^n\int_{-\infty}^{\infty}dk_z\vct{P}^{\textrm{out}}_{n,k_z}(\vct{r})\cdot\vct{P}^{'\textrm{out}}_{-n,-k_z}(\vct{r})T_{n,k_z}^{PP'},
\label{eq:GT_trace_expansion}
\end{equation}
or from Eqs.~\eqref{eq:GTParallel} and~\eqref{eq:GT11Parallel}--\eqref{eq:GT33Parallel} by letting $ d \to 0 $:
\begin{align}
\notag \Tr\mathbb{G}_{\mathbb{T}}(\vct{r},\vct{r}) = \ & \frac{i}{4\pi}\int_0^{\infty}dk_z\left\{H_1^2(qr)T_{0,k_z}^{MM}+\left[\frac{k_z^2}{k^2}H_1^2(qr)+\frac{q^2}{k^2}H_0^2(qr)\right]T_{0,k_z}^{NN}\right\}\\
\notag & + \frac{i}{2\pi}\sum_{n=1}^{\infty}\int_0^{\infty}dk_z\Bigg\{\left[\frac{n^2}{(qr)^2}H_n^2(qr)+\left[H'_n(qr)\right]^2\right]T_{n,k_z}^{MM} + 4\frac{nk_z}{kqr}H_n(qr)H'_n(qr)T_{n,k_z}^{MN}\\
& \ \ \ \ \ \ \ \ \ \ \ \ \ \ \ \ \ \ \ \ \ \ \ \ \ + \left[\frac{k_z^2}{k^2}\left[H'_n(qr)\right]^2+\frac{n^2k_z^2}{k^2(qr)^2}H_n^2(qr)+\frac{q^2}{k^2}H_n^2(qr)\right]T_{n,k_z}^{NN}\Bigg\}.
\label{eq:GT_trace}
\end{align}
Then one takes the imaginary part of Eq.~\eqref{eq:GT_trace}. As expected from physical grounds, the trace in Eq.~\eqref{eq:GT_trace} does not depend on $ \varphi $ and $ z $.

For a perfectly conducting cylinder, Eq.~\eqref{eq:GT_trace} simplifies to
\begin{align}
\notag & \lim_{|\varepsilon|\to\infty} \Tr\mathbb{G}_{\mathbb{T}}(\vct{r},\vct{r}) = -\frac{i}{4\pi}\int_0^{\infty}dk_z\left\{H_1^2(qr)\frac{J_1(qR)}{H_1(qR)}+\left[\frac{k_z^2}{k^2}H_1^2(qr)+\frac{q^2}{k^2}H_0^2(qr)\right]\frac{J_0(qR)}{H_0(qR)}\right\}\\
& - \frac{i}{2\pi}\sum_{n=1}^{\infty}\int_0^{\infty}dk_z\Bigg\{\left[\frac{n^2}{(qr)^2}H_n^2(qr)+\left[H'_n(qr)\right]^2\right]\frac{J'_n(qR)}{H'_n(qR)} + \left[\frac{k_z^2}{k^2}\left[H'_n(qr)\right]^2+\frac{n^2k_z^2}{k^2(qr)^2}H_n^2(qr)+\frac{q^2}{k^2}H_n^2(qr)\right]\frac{J_n(qR)}{H_n(qR)}\Bigg\}.
\label{eq:GT_trace_mirror}
\end{align}
It can be shown that, for the imaginary part of the trace in Eq.~\eqref{eq:GT_trace_mirror}, the integration can be restricted to $ k_z \leq k $:
\begin{align}
\notag & \Im \lim_{|\varepsilon|\to\infty} \Tr\mathbb{G}_{\mathbb{T}}(\vct{r},\vct{r}) = -\frac{1}{4\pi}\int_0^{k}dk_z\Re\left\{H_1^2(qr)\frac{J_1(qR)}{H_1(qR)}+\left[\frac{k_z^2}{k^2}H_1^2(qr)+\frac{q^2}{k^2}H_0^2(qr)\right]\frac{J_0(qR)}{H_0(qR)}\right\}\\
& - \frac{1}{2\pi}\sum_{n=1}^{\infty}\int_0^{k}dk_z\Re\Bigg\{\left[\frac{n^2}{(qr)^2}H_n^2(qr)+\left[H'_n(qr)\right]^2\right]\frac{J'_n(qR)}{H'_n(qR)} + \left[\frac{k_z^2}{k^2}\left[H'_n(qr)\right]^2+\frac{n^2k_z^2}{k^2(qr)^2}H_n^2(qr)+\frac{q^2}{k^2}H_n^2(qr)\right]\frac{J_n(qR)}{H_n(qR)}\Bigg\}.
\label{eq:GT_ImTr_mirror}
\end{align}

% APPENDIX_end ------------------------------------------

\twocolumngrid
% BIBLIOGRAPHY_begin ------------------------------------------

%apsrev4-2.bst 2019-01-14 (MD) hand-edited version of apsrev4-1.bst
%Control: key (0)
%Control: author (72) initials jnrlst
%Control: editor formatted (1) identically to author
%Control: production of article title (-1) disabled
%Control: page (0) single
%Control: year (1) truncated
%Control: production of eprint (0) enabled
%

% BIBLIOGRAPHY_end ------------------------------------------

\end{document}